\documentclass[12pt]{article}
\usepackage{savesym}
\usepackage{cite}
\usepackage{wasysym}
\usepackage{amscd}
\usepackage{graphicx}
\usepackage{pifont}
\usepackage{float}
\usepackage{subfig}
\usepackage[all]{xy}
\usepackage{multicol}
\usepackage{amsfonts}
\usepackage{picture}
\usepackage{amssymb}
\savesymbol{iint}
\savesymbol{iiint}
\usepackage{amsmath}
\restoresymbol{TXF}{iint}
\restoresymbol{TXF}{iiint}
\usepackage{color}
\usepackage{clay}
\usepackage{hyperref}
\usepackage{slashed}
\hypersetup{colorlinks=true}
\hypersetup{linkcolor=black}
\hypersetup{citecolor=black}
\hypersetup{urlcolor=black}
\usepackage{setspace}
\usepackage{multirow}
\usepackage{wrapfig}
\usepackage{verbatim}
\numberwithin{equation}{section}

\begin{document}
\date{May, 2016}

\institution{Fellows}{\centerline{${}^{1}$School of Natural Sciences, Institute for Advanced Study, Princeton, NJ, USA}}
\institution{HarvardU}{\centerline{${}^{2}$Jefferson Physical Laboratory, Harvard University, Cambridge, MA, USA}}

\title{Toda Theory From Six Dimensions}

\authors{Clay C\'{o}rdova,\worksat{\Fellows}\footnote{e-mail: {\tt claycordova@ias.edu}}  and Daniel L. Jafferis \worksat{\HarvardU}\footnote{e-mail: {\tt jafferis@physics.harvard.edu}} }

\abstract{We describe a compactification of the six-dimensional (2,0) theory on a four-sphere which gives rise to a two-dimensional Toda theory at long distances.  This construction realizes chiral Toda fields as edge modes trapped near the poles of the sphere.    We relate our setup to compactifications of the $(2,0)$ theory on the five and six-sphere.  In this way, we explain a connection between half-BPS operators of the $(2,0)$ theory and two-dimensional W-algebras, and derive an equality between their conformal anomalies.  As we explain, all such relationships between the six-dimensional (2,0) theory and Toda field theory can be interpreted as statements about the edge modes of complex Chern-Simons on various three-manifolds with boundary. }

\maketitle

\setcounter{tocdepth}{2}
\tableofcontents
\section{Introduction}

\label{intro}

In this paper we construct a compactification of the six-dimensional $(2,0)$ conformal field theory on $S^{4}$.  We demonstrate that the resulting low-energy effective action in two dimensions is a complexified Toda field theory.   Thus, we obtain a direct derivation of the Alday-Gaiotto-Tachikawa correspondence \cite{Alday:2009aq, Wyllard:2009hg, Alday:2009fs}. Previous approaches to this relationship have appeared in \cite{Dijkgraaf:2009pc, Nekrasov:2010ka, Mironov:2010pi, Alfimov:2011ju, Yagi:2012xa, Vartanov:2013ima, Aganagic:2013tta, Mironov:2015thk, Nekrasov:2015wsu}.  We further apply our construction to relate various properties of the Toda theory, such as a its spectrum of chiral operators and associated central charge, to limits of the six-dimensional superconformal index and $c$-type Weyl anomaly as anticipated by the results of \cite{Alday:2009qq, Beem:2014kka}.

\subsection{Context of the AGT Conjecture}

The six-dimensional $(2,0)$ theories \cite{Witten:1995zh, Strominger:1995ac, Witten:1995em} are maximally supersymmetric and conformally invariant.  They are labelled by an ADE Lie algebra $\frak{g}$.  See for instance \cite{moorefklect,Cordova:2015vwa, Beem:2015aoa} for a recent survey of their general properties from various perspectives.  Starting from these theories, a large class of lower-dimensional supersymmetric theories may be constructed by twisted compactification on manifolds of various dimensions \cite{Witten:1997sc, Gaiotto:2009we, Gaiotto:2009hg, Dimofte:2011ju, Cecotti:2011iy, Gadde:2013sca}.  A widely studied example results in four-dimensional $\mathcal{N}=2$ theories.  These theories are labelled ${\cal T}_{\frak{g}}(\Sigma)$, where $\Sigma$ is the compactification Reimann surface (possibly with punctures).

This geometric perspective on four-dimensional field theories yields insight into many of their physical properties such as non-trivial dualities \cite{Gaiotto:2009we}, moduli spaces \cite{Witten:1997sc}, BPS particles \cite{Klemm:1996bj, Gaiotto:2009hg, Alim:2011ae}, and spectrum of local operators \cite{Gadde:2009kb, Gadde:2011ik}.  The AGT correspondence studied here is a dramatic example in this vein.  In this case, the four-dimensional observable of interest is the $S^{4}$ partition function computed in \cite{Pestun:2007rz, Hama:2012bg}.  This partition function takes the form of an integral over Coulomb branch parameters $u$.  Schematically, for any gauge theory one has
\begin{equation}
\mathcal{Z}_{S^{4}}=\int du~ e^{-S(u)} |\mathcal{F}(u)|^{2}~,
\end{equation}
where $\mathcal{F}(u)$ are generating functions of pointlike instantons \cite{Nekrasov:2002qd} localized at the two poles of the sphere.

In the context of the theories ${\cal T}_{\frak{g}}(\Sigma)$ this partition function is reinterpreted as a two-dimensional observable on $\Sigma$.  The factors $\mathcal{F}(u)$ are conformal blocks for a Toda field theory on $\Sigma$, where $u$ plays the role of the exchanged operator dimension, and external vertex operators arise from codimension two defects of the six-dimensional theory which puncture the sphere.  The integration over $u$ produces a Toda correlation function.

The six-dimensional origin of the theories ${\cal T}_{\frak{g}}(\Sigma)$ suggests a natural explanation for the correspondence: \emph{the six-dimensional $(2,0)$ theory reduced on $S^{4}$ is the Toda field theory}.  Then, the correspondence of observables arises from a commuting the order of compactification on $\Sigma$ and $S^{4}$.  Our main result is to carry out the dimensional reduction on $S^{4}$ and explain how Toda theory emerges.

In fact, in our construction we will naturally encounter a complexified Toda field theory.  The dynamical variables consist of $r$ complex bosons $\Phi$ where $r$ is the rank of $\mathfrak{g}$.  The action takes the form
\begin{equation}
S=\frac{q}{8\pi} \int dz d\bar{z}~C_{ij}\partial\Phi^{i}\bar{\partial}\Phi^{j}+\sum_{i}\exp\left(C_{ij}\Phi_{j}\right) +\frac{\tilde{q}}{8\pi} \int dz d\bar{z}~C_{ij}\partial\bar{\Phi}^{i}\bar{\partial}\bar{\Phi}^{j}+\sum_{i}\exp\left(C_{ij}\bar{\Phi}_{j}\right) ~,\label{complexToda}
\end{equation}
where $C_{ij}$ is the Cartan matrix of $\frak{g}$ and the coupling constant $q$ is expressed as
\begin{equation}
q=k+is~, \hspace{.5in}\tilde{q}=k-is~.
\end{equation}
The parameters $k$ and $s$ are determined by the geometry of our compactification.  Specifically, $k$  which is an integer, occurs when we include the orbifold singularity $S^{4}\rightarrow S^{4}/\mathbb{Z}_{k},$ and $s$ (which is either real or pure imaginary) may be activated by deforming the metric away from the round geometry.  The special case of the round $S^{4}$ which is the subject of the original AGT conjecture \cite{Alday:2009aq, Wyllard:2009hg, Alday:2009fs} has $k=1$ and $s=0$.

It is worth observing that the original form of the AGT correspondence relates to real Toda theory, not the complexified version we arrive at.  The relation between the real and complex theories involves a duality.  Specifically, in \cite{Nishioka:2011jk} it was suggested (based in part on evidence from \cite{Belavin:2011pp}) that the $\frak{su}(n)$ $(2,0)$ theory on $S_{\ell}^{4}/\mathbb{Z}_{k}$ can be described in terms of a real parafermionic Toda theory and a decoupled coset model.  The dynamical variables of the paraToda theory consist of $n-1$ parafermions $ \psi_{i}$ and $n-1$ real bosons $\phi_{j}$ . The action takes the form \cite{LeClair:1992xi}
\begin{equation}
S_\textrm{para-Toda}= S \left(\frac{\frak{\hat{su}(n)}_{k}}{\hat{\frak{u}}(1)^{n-1}}\right)+\int dxdy \left[\partial_{\mu}\phi\partial_{\mu} \phi+ \sum_{i=1}^{n-1} \psi_{i} \bar{\psi}_{i}\exp\left(\frac{b}{\sqrt{k}}C_{ij} \phi_{j} \right)\right]. \label{paratoda}
\end{equation}
In the above, the first term describes the coset model parafermions $\psi$ associated the simply connected form of $\mathfrak{su}(n)$  at level $k$ \cite{Gepner:1987sm} and $b$ is a coupling constant.  The parafermion operator $\psi_{i} \bar{\psi}_{i}$ has left and right scaling dimension $1-1/k,$ and the background charge of the bosons $\phi_{j}$ is adjusted so that the interaction is marginal for any $b$.  Comparing to our result implies the duality
\begin{equation}
\mathrm{complex~ Toda}(n,k,s)\leftrightarrow \mathrm{real~paraToda}(n,k,b)+\frac{\hat{\frak{su}}(k)_{n}}{\hat{\frak{u}}(1)^{k-1}} ~,\label{duality}
\end{equation}
where the two models on the right-hand-side of the above are decoupled, and the map between parameters is
\begin{equation}
b=\sqrt{\frac{k-is}{k+is}}~.\label{bintro}
\end{equation}
Note in particular that in the special case $k=1$ the parafermions and decoupled coset have no degrees of freedom and the right-hand-side of \eqref{duality} is the ordinary Toda theory.

We conjecture that this duality holds for all values of the parameters. In the special case $n=2$ and $k=1$ this duality has been checked by a direct examination of the relevant Hilbert spaces \cite{Dimofte:2014zga}.  Meanwhile for $n=k=2$ the above is a relation between complex Liouville theory, and real superLiouville theory plus a decoupled fermion and is equivalent to a kind of bosonization  \cite{Belavin:2011sw, Schomerus:2012se}.  It is an interesting problem to establish \eqref{duality} in generality.

\subsection{Overview of the Derivation}

We begin with the configuration in six dimensions. Using background off-shell supergravity methods, we obtain a supersymmetric theory on the Riemannian product $S_{\ell}^{4}/\mathbb{Z}_{k}\times \Sigma$ where $\ell$ is a real squashing parameter.  A significant feature of this background is that the two factors in the geometry are treated in an asymmetric fashion.  Along the Riemann surface directions the theory is topologically twisted, and a non-trivial spin connection on $\Sigma$ is compensated for by an $so(2)_{R}$ gauge field.  This implies that in the limit of a large $S^{4}$ our construction will give rise to the effective four-dimensional theory $\mathcal{T}_{\frak{g}}(\Sigma)$.  By contrast, along the $S^{4}$ directions the theory is not twisted, and the background is of the type studied in \cite{Pestun:2007rz}.

For this reason, the six-dimensional background fields do not factorize into a decoupled product of four-dimensional and two-dimensional pieces, even though the metric does. This turns out to be crucial to understanding the most surprising aspects of the correspondence - namely that the effective theory on $\Sigma,$ the Toda field theory, is neither supersymmetric, nor a standard gauge theory.

The resolution of this puzzle is that the even in the limit where $\Sigma$ is taken to be large, and hence replaced by $\mathbb{R}^{2},$ the theory is already effectively topologically twisted along these directions.   One way to see this is to first place the $(2,0)$ SCFT on the conformally flat space $H_2 \times S^4$, where $H_{2}$ is the hyperbolic plane and the two factors have equal radii. Then $H_2$ may be topologically twisted to flat $\mathbb{R}^2$. This results in the 6d background relevant for AGT in the case of the round $S^4$, and makes clear that the supergravity background fields are non-trivially activated even in flat $\mathbb{R}^2$.

From this description of the background we can also understand a key point: the commutativity of the compactifications on $\Sigma$ and $S^{4}$.  The six-dimensional theory is Weyl invariant.  Therefore (up to an overall factor due to the Weyl anomaly) the partition function is independent of the overall scale of the six-manifold, $S_{\ell}^{4}/\mathbb{Z}_{k} \times \Sigma$.  Moreover, the topological twisting on $\Sigma$ implies that changing the volume of $\Sigma$ is separately a $Q$-exact deformation. Therefore the supersymmetric partition function is separately independent of both sizes.  We thus focus on the limit of small $S_{\ell}^{4}/\mathbb{Z}_{k}$ and describe the effective two-dimensional field theory that emerges.

If there were a known Lagrangian description of the (2,0) theories, it would be straightforward to perform such a reduction.  Of course, the lack of such a Lagrangian is part of the reason why these theories and their four-dimensional children are interesting.  Instead of proceeding directly, we will derive the two-dimensional theory using a combination of input from the circle reduction of the $(2,0)$ theory, five-dimensional Yang-Mills theory, and our previous investigation of the $(2,0)$ theory on backgrounds of the form $M_{3}\times S^{3}_{\ell}$ where the first factor is a general three manifold and the second factor is a squashed $S^{3}$.

Backgrounds of the form $M_{3}\times S^{3}_{\ell}/\mathbb{Z}_{k}$ are the subject of the $3d$-$3d$ correspondence \cite{Dimofte:2011ju, Dimofte:2011py}.  This three-dimensional relationship is similar in spirit to the AGT conjecture.  It expresses the $S^{3}$ partition function of a three-dimensional theory $\mathcal{T}_{\frak{g}}(M_{3}),$ obtained by compactification of the $(2,0)$ theory on a three-manifold $M_{3}$, in terms of the Chern-Simons partition function with complexified gauge algebra $\frak{g}_{\mathbb{C}}$ on the three-manifold $M_{3}$.  This $3d$-$3d$ relationship has been derived in \cite{Yagi:2013fda, Cordova:2013cea, Lee:2013ida, Dimofte:2014zga} by demonstrating that the $(2,0)$ theory compactified on $S^{3}_{\ell}/\mathbb{Z}_{k}$ gives rise to complex Chern-Simons theory at long distances.  Here, the geometric parameters $k$ and $\ell$ map on to coupling constants, the complex level, of the Chern-Simons theory.\footnote{Recently, a $4d$-$2d$ correspondence, related to compactifications of the $6d$ theory on four-manifolds, has been investigated in \cite{Assel:2016lad}.}

\begin{figure}
  \centering
\includegraphics[width=0.5\textwidth]{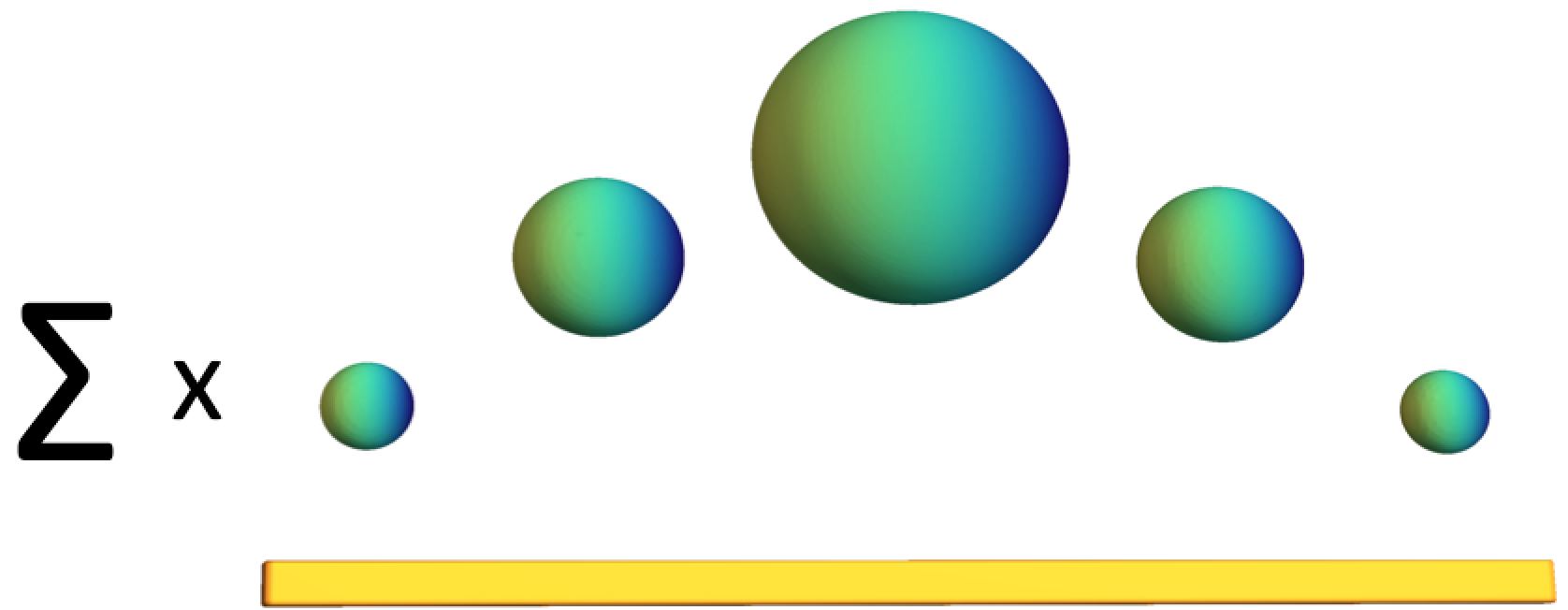}
  \caption{The six-dimensional geometry relevant to the AGT conjecture.  The $S^{4}$ is viewed as a fibration of $S^{3}$ over a line.  By reducing on $S^{3}$ we obtain complex Chern-Simons theory on the product of the Riemann surface $\Sigma$ times an interval. }
  \label{fig1}
\end{figure}

To leverage the $3d$-$3d$ correspondence in the present context, we must find a suitable $S^{3}_{\ell}/\mathbb{Z}_{k}$ hiding inside our geometry.  We can achieve this by using the Weyl invariance of the $(2,0)$ conformal field theory.  The four-dimensional space $S^{4}_{\ell}/\mathbb{Z}_{k}$ in our construction may be viewed as an $S^{3}$ fibered over an interval, which collapses at the two ends.  Up to a conformal transformation this is simply $S^{3}_{\ell}/\mathbb{Z}_{k}$ times a line.  Therefore as an intermediate step, we may reduce to three-dimensions resulting in complex Chern-Simons gauge theory on the warped product $\Sigma \times \mathbb{R}$.  This logic is illustrated in Figure \ref{fig1}.

To pass from these considerations to a  two-dimensional effective description we must carefully keep track of the boundary data, now at infinity in $\Sigma \times \mathbb{R}.$  These boundary conditions arise from the fact that, before the Weyl transformation, the equatorial $S^{3}$ collapses at the boundaries of the line.  Phrased in this language the AGT correspondence means that the theory of edge modes, obtained by enforcing appropriate boundary conditions on complex Chern-Simons theory, is the Toda field theory.  This perspective on AGT is similar to the point of view of \cite{Nekrasov:2010ka, Yagi:2012xa, Vafa:2015euh}. 

One way to understand the boundary conditions is to view the situation from the point of view of $5d$ super Yang-Mills.  Locally, near the ends of the interval where $S^{3}_{\ell}/\mathbb{Z}_{k}$ collapses the geometry is that of (an orbifold of) Taub-NUT space.  Therefore, reducing along the Hopf fiber of $S^{3}_{\ell}/\mathbb{Z}_{k},$ the boundary condition in question is similar to D4-branes ending on D6-branes.  This is the Nahm pole boundary condition studied in \cite{Diaconescu:1996rk, Gaiotto:2008sa, Gaiotto:2008ak }.  Near the boundary (locally at $\sigma=0$) a triplet of scalar fields in the Yang-Mills multiplet have a simple pole and obey
\begin{equation}
\frac{dX_{i}}{d\sigma}+\varepsilon_{ijk}X_{j}X_{k}=0~.
\end{equation}
The resulting brane configuration is a non-commutative funnel in which the D4 branes merge into the D6 brane.

Our final step is therefore to translate the Nahm pole boundary conditions into a Chern-Simons language.  Near the boundary, the Chern-Simons gauge field approaches a pure gauge transformation and can therefore be described by a holomorphic current $J$ in a boundary WZW model based on the algebra $\frak{g}_{\mathbb{C}}$.  The Nahm pole boundary condition is conveniently phrased as a constraint on these currents and reduces them to a chiral complex Toda theory.   The fusion of two boundaries thus gives rise to the full non-chiral system described by the Lagrangian \eqref{complexToda}.  This is a simple complexification of well-known relations between the real Toda theory and the $\frak{sl}(n,\mathbb{R})$ WZW model \cite{Forgacs:1989ac, Balog:1990mu}.

Finally in section \ref{sec:duality} we discuss the duality between complex Toda theory and parafermionic toda plus a coset model.  We motivate the identification of parameters \eqref{bintro}, and give some further evidence from the more  detailed analysis \cite{Dimofte:2014zga, Belavin:2011sw, Schomerus:2012se}. 

\subsection{Further Applications}

One of the virtues of deriving the AGT conjecture directly from an action formalism is that we can readily see how other known relationships between the Toda field theory and the $(2,0)$ theory emerge from the same basic logic.  In section \ref{sec:operators} we focus in particular on the half-BPS local operators of the $(2,0)$ theory.

Recently in \cite{Beem:2014kka} it has was observed that these operators are in one-to-one correspondence with the chiral Toda operators and that correlators of these operators, when restricted to a two-plane, reproduce exactly the chiral Toda correlation functions.  In particular, the $c$-type Weyl anomaly of the six-dimensional $(2,0)$ theory matches that of chiral Toda at $b=1$, as first observed in \cite{Alday:2009qq}.

To make contact with our previous work, it is fruitful to rephrase these statements as properties of partition functions of the $(2,0)$ theory.  The fact that the set of half-BPS operators coincide with the chiral Toda currents means that a limit of the $(2,0)$ superconformal index, or the partition function on $S^{1}\times S^{5},$ is equal to the chiral Toda partition function on a torus.  This correspondence is natural from a Chern-Simons edge mode point of view.  We view the $S^{1}\times S^{5}$ as an $S^{3}$ fibered over a solid torus where the $S^{3}$ shrinks at the boundary.  Reducing on this $S^{3}$ yields complex Chern-Simons on the solid torus, but now the edge modes are chiral since there is a single boundary component of the three-manifold.  Applying the same logic as before yields the desired relation between the $(2,0)$ index and the chiral Toda partition function.

Parallel analysis may be applied to correlators.  In this case we view the flat space correlators of the $(2,0)$ theory as conformally equivalent to correlators on an $S^{6}$.  Then we view $S^{6}$ as an $S^{3}$ fibered over a solid ball.  Reducing on the $S^{3}$ as above and passing to edge modes, we obtain chiral Toda correlators on the boundary of the ball, which lifts to a two-pane in six-dimensions.

\begin{figure}
  \centering
  \subfloat[]{\label{fig:1}\includegraphics[width=0.3\textwidth, height=0.25\textwidth]{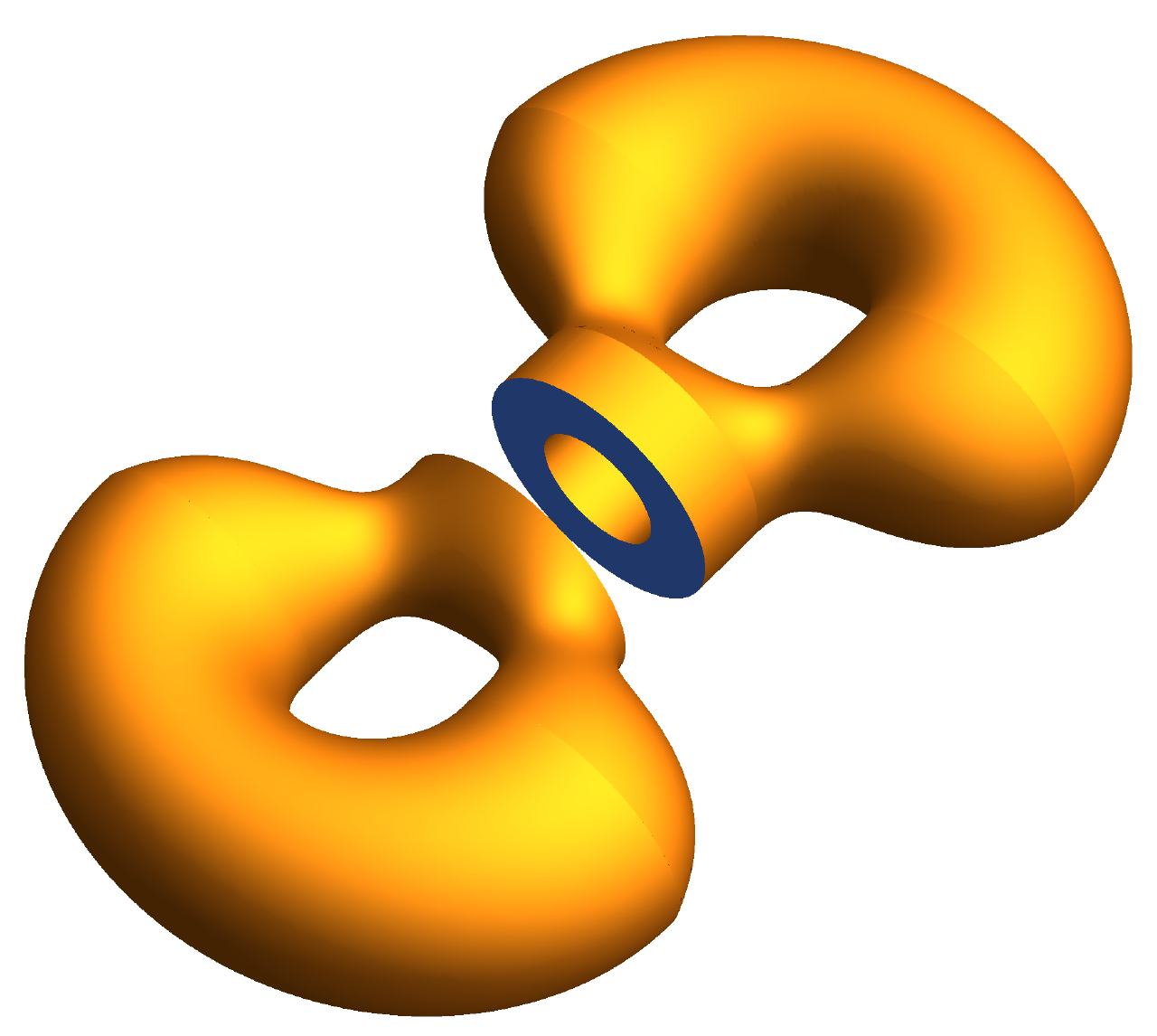}}
  \hspace{.4in}
  \subfloat[]{\label{fig:2}\includegraphics[width=0.25\textwidth, height=0.25\textwidth]{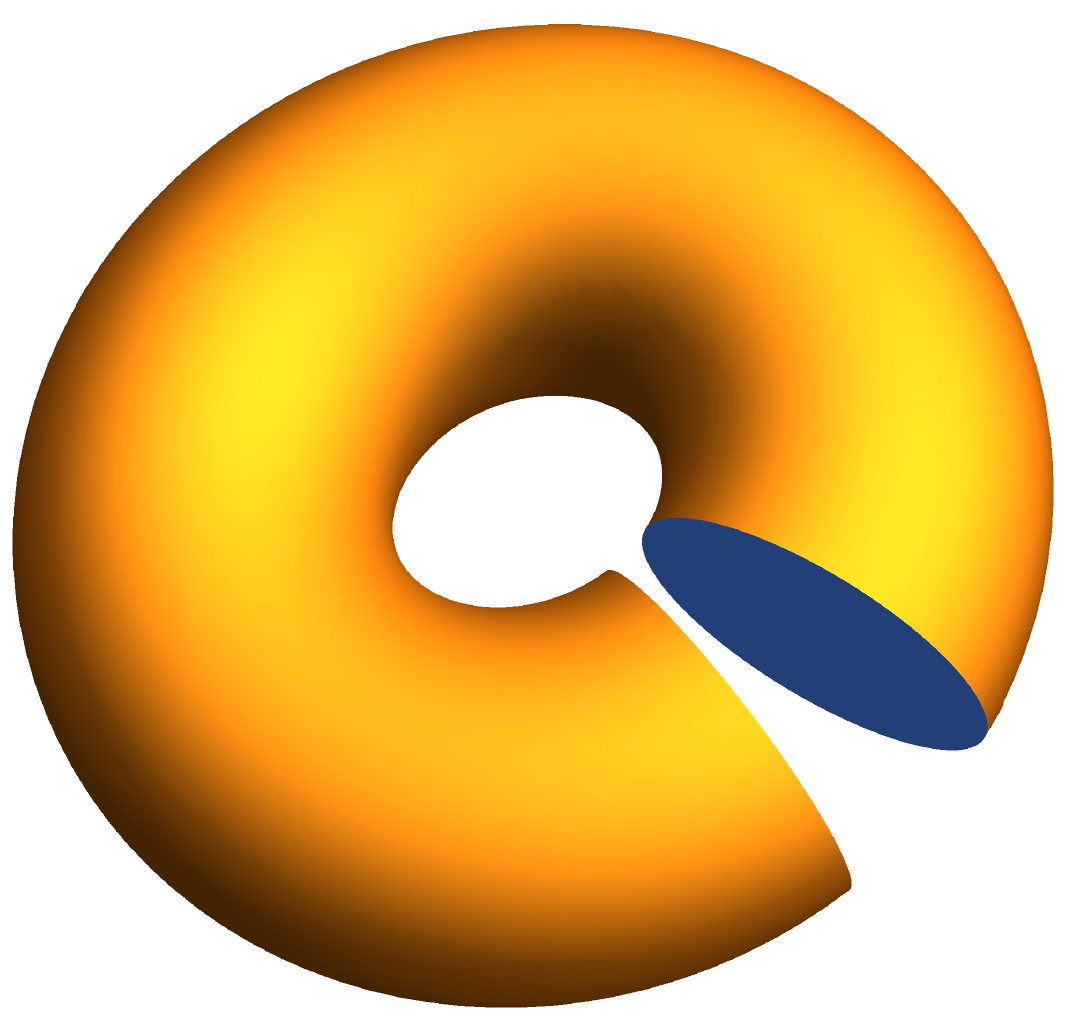}}
  \hspace{.4in}
    \subfloat[]{\label{fig:3}\includegraphics[width=0.25\textwidth, height=0.25\textwidth]{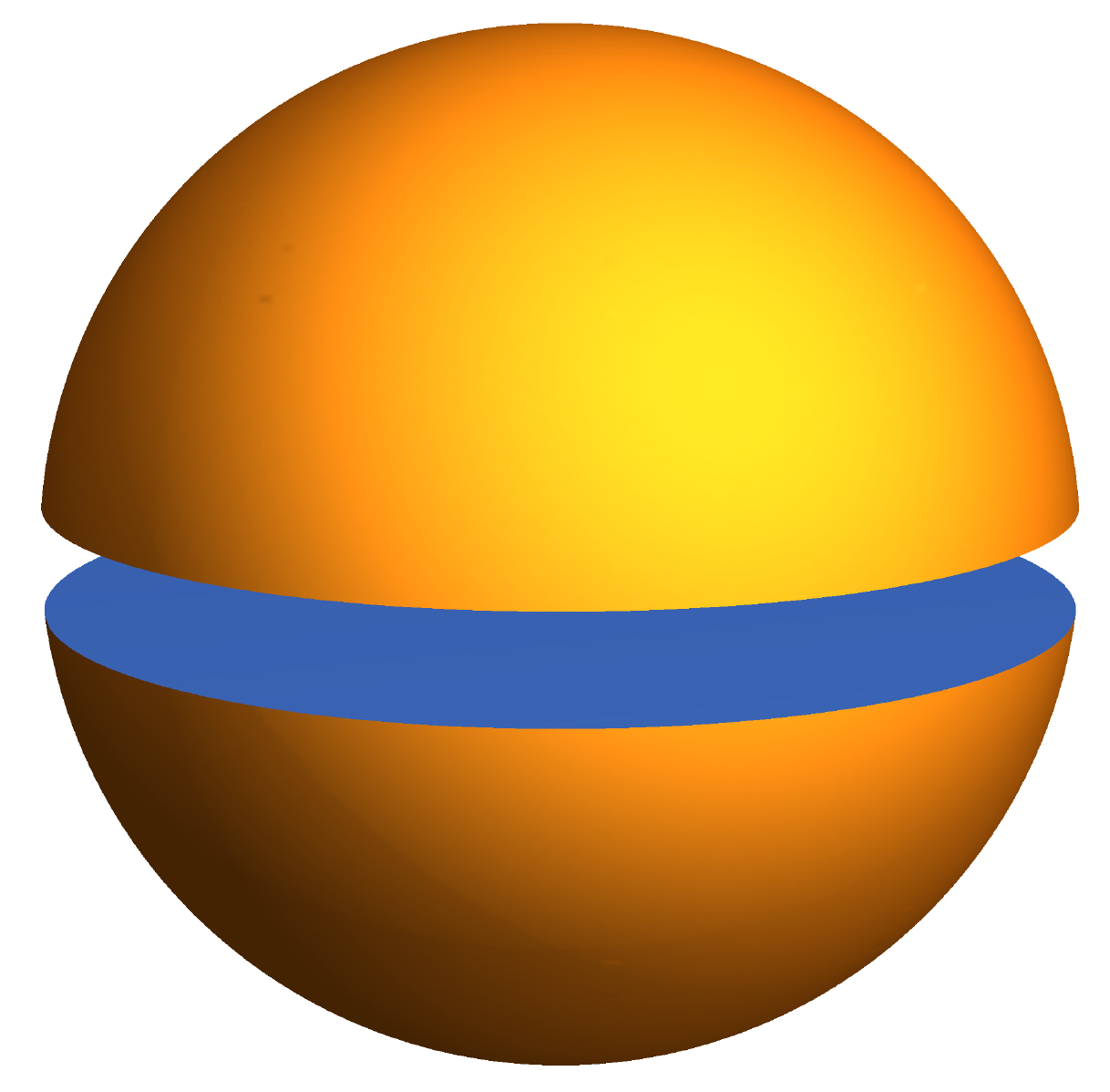}}
  \caption{By reducing on $S^{3}$'s, various relations between the $(2,0)$ theory and Toda arise from complex Chern-Simons theory on manifolds with boundary (shown above with sections removed to illustrate thickness).   In (a), a thickened Riemann surface $\Sigma$ has non-chiral edge modes and yields the relationship between the $S^{4}$ partition function and the non-chiral Toda correlator.  In (b), reducing on an $S^{3}$ relates the geometry relevant for the 6$d$ superconformal index to complex Chern-Simons on the solid torus and hence a chiral Toda partition function.  In (c) the $S^{6}$ partition function of the $(2,0)$ theory gives rise to complex Chern-Simons on the solid ball, and hence chiral Toda correlators. }
  \label{fig}
\end{figure}

Thus we see that various relationships between the $(2,0)$ theory and the Toda theory may be fruitfully encoded by complex Chern-Simons on manifolds with boundary as illustrated in Figure \ref{fig}.

\section{The $(2,0)$ Theory on a Squashed $S^{4}$}

In this section we construct a Euclidean compactification of the six-dimensional $(2,0)$ theory on a four-dimensional spherical background.  Our construction yields supersymmetric compactifications on geometries of the form $S^{4}_{\ell}/\mathbb{Z}_{k}\times \Sigma$ where  $\Sigma$ is any Riemann surface (possibly with punctures), and $S^{4}_{\ell}/\mathbb{Z}_{k}$ is a geometrically squashed version of the orbifold $S^{4}/\mathbb{Z}_{k}$ discussed below.  Ultimately, we will see that $\ell$ and $k$ are related to the parameters of the resulting Toda system.  The guiding principle governing our construction is that it must preserve supersymmetry on any Riemann surface $\Sigma$ independent of the choice of metric.  Therefore the background must be topologically twisted on $\Sigma$.  By contrast supersymmetry requires a constrained form of the metric on $S^{4}_{\ell}/\mathbb{Z}_{k}.$

In the limit where the radius of $S^{4}_{\ell}/\mathbb{Z}_{k}$ is taken to infinity (so that it may be replaced by $\mathbb{R}^{4}$), our background will reduce to the ordinary toplogically twisted compactification of the $(2,0)$ theory on a Riemann surface which gives rise to the four-dimensional field theory $T_{\frak{g}}(\Sigma)$.  Viewed in terms of symmetries, the six-dimensional Lorentz and R-symmetry group is $so(6)_{L}\times so(5)_{R}$.  In the limit of infinite radius of $S^{4}_{\ell}/\mathbb{Z}_{k},$ the pattern of symmetry breaking may be understood as
\begin{equation}
so(6)_{L}\times so(5)_{R} \rightarrow so(4)_{L}\times so(2)_{l}\times so(3)_{R} \times so(2)_{r} \rightarrow so(4)_{L}\times so(3)_{R}\times so(2)_{\Delta},
\label{symmetrybreaking}
\end{equation}
where $so(4)_{L}$ is the $\mathbb{R}^{4}$ Lorentz group, $so(3)_{R}$ is an $R$-symmetry of the four-dimensional $\mathcal{N}=2$ theory, and $so(2)_{\Delta}$ is the diagonal subgroup $so(2)_{r}\times so(2)_{l}.$

In generalizing away from the infinite radius limit of $S^{4}_{\ell}/\mathbb{Z}_{k}$ to the full background of interest, the $so(4)_{L}$ symmetry will be partially broken by the geometry, and the $so(3)_{R}$ symmetry will be broken down to a $so(2)_{R}$ subgroup.  However, the $so(2)_{\Delta}$ must remain unbroken to ensure compatibility with twisting on $\Sigma$.

In general, it is convenient to phrase the requirements of supersymmetry using the coupling to background supergravity source fields \cite{Buchbinder:1998qv, Festuccia:2011ws}.  Aspects of the relevant supergravity multiplets have been described in \cite{Bergshoeff:1999db, Cordova:2013bea}.  Among the fields, are the metric and an $so(5)_{R}$ gauge field.  Topological twisting along $\Sigma$ means correlating some components of the $R$-gauge field with the spin connection.  Besides the metric and $R$-gauge field, there are other bosonic fields in the supergravity multiplet and we will be forced to activate them.  The complete list of supergravity source fields and their coupling to the $(2,0)$ theory are dictated by the $6d$ (2,0) energy-momentum tensor supermultiplet, and are enumerated in Table \ref{table6dfields}.
\begin{table}[h]
\centering
\begin{tabular}{|c|c|c|c|}
\hline
Field & Type & $so(5)_{R} $  \\
\hline
 \multirow{2}{*}{$g$} & \multirow{2}{*}{Metric }  &\multirow{2}{*}{$\mathbf{1}$} \\
 & &    \\
\hline
 \multirow{2}{*}{$V$} & \multirow{2}{*}{$R$-gauge field }  &\multirow{2}{*}{$\mathbf{10}$} \\
 & &    \\
\hline
\multirow{2}{*}{$ T$} & \multirow{2}{*}{anti-self-dual three-form} & \multirow{2}{*}{$\mathbf{5}$}  \\
 & &  \\
\hline
\multirow{2}{*}{$ D$} & \multirow{2}{*}{scalar } & \multirow{2}{*}{$\mathbf{14}$} \\
 & &  \\
\hline
\end{tabular}
\caption{Bosonic fields of six-dimensional $(2,0)$ off-shell supergravity.}
\label{table6dfields}
\end{table}
A configuration of source fields is supersymmetric if the variation of the supergravity fermions vanishes.

We now proceed to describe the background in detail.  We begin with the metric.  As advertised, this is a squashed version of the orbifold $S^{4}/\mathbb{Z}_{k}\times \Sigma$
\begin{equation}
ds^{2}  =   r^{2}\left[d\sigma^{2}+\frac{f(\sigma)^{2}\ell^{2}}{4} \left(d\theta^{2}+\sin^{2}(\theta)d\phi^{2}\right) +\frac{f(\sigma)^{2}}{k^{2}}\left(d\psi+k\cos^{2}(\theta/2)d\phi\right)^{2}\right] + r^{2}e^{A(x,y)}(dx^{2}+dy^{2}). \label{metric}
\end{equation}
In the above, the coordinates  $(\theta, \phi, \sigma, \psi)$ have the following ranges
\begin{equation}
0\leq \theta \leq \pi~, \hspace{.5in} 0\leq \phi \leq 2\pi~, \hspace{.5in} 0\leq \sigma \leq \pi~, \hspace{.5in} 0\leq \psi \leq 2\pi~.
\end{equation}
The parameter $\ell$ is a real squashing deformation of the round metric.  It is related to $s$ as
\begin{equation}
s=\sqrt{\ell^{2}-1}~.
\end{equation}
Meanwhile, $x,y$ are local coordinates on the Reimann surface $\Sigma,$ and $A(x,y)$ is a local conformal factor of the two-dimensional metric.

The function $f(\sigma)$ appearing above is constrained to have simple zeros at $\sigma=0, \pi,$ and to be positive definite on the interval away from these loci.  Aside from these conditions $f(\sigma)$ may be chosen arbitrarily and its precise form will not effect our construction.  The special case of the round orbifold $S^{4}/\mathbb{Z}_{k}$ is achieved when $\ell=1$ and $f(\sigma)=\sin(\sigma).$  One may view the the four-dimensional metric on $S^{4}_{\ell}/\mathbb{Z}_{k}$ as a fibration of a Lens space $S^{3}/\mathbb{Z}_{k}$ over the interval with parameter $\sigma$.  At the ends of the interval, the lens space collapses.

While the exact form of $f(\sigma)$ will not enter our analysis, it is worthwhile to note that the class of metric backgrounds we are investigating have curvature singularities. Let us focus on the four-dimensional metric on $S^{4}_{\ell}/\mathbb{Z}_{k}$ in the region near $\sigma=0$ where the equatorial space collapses.  We parameterize $f(\sigma)$ in a Taylor series as
\begin{equation}
f(\sigma)=\alpha \sigma +\frac{\beta}{2}\sigma ^{2}+\mathcal{O}(\sigma^{3})~.
\end{equation}
Then one finds the values of various curvature invariants to be
\begin{eqnarray}
R & = & \left(\frac{8\ell^{2}-6\alpha^{2}\ell^{4}-2}{\alpha^{2} r^{2}\ell^{4}}\right)\left(\frac{1}{\sigma^{2}}\right)+\left(\frac{2\beta-8\ell^{2}\beta-12\beta \alpha^{2}\ell^{4}}{\alpha^{3}r^{2}\ell^{4}}\right)\left(\frac{1}{\sigma}\right)+\mathcal{O}(\sigma^{0})~, \nonumber \\
R_{\mu\nu}R^{\mu\nu} & = & \left(\frac{12 - 32 \ell^2 + 32 \ell^4 + 8 \alpha^2 \ell^4 - 32 \alpha^2 \ell^6 + 12 \alpha^4 \ell^8}{\alpha^4 \ell^8 r^4 }\right)\left(\frac{1}{\sigma^{4}}\right)+\mathcal{O}(\sigma^{-3})~.
\end{eqnarray}
Except for the special case of the round background $(\ell=1)$ it is not possible to eliminate these curvature singularities by a choice of $\alpha$ and $\beta$.  We can however choose $\alpha$ and $\beta$ such that the Ricci scalar singularities vanish.  In that case singular scalar curvature invariants only occur from contraction of more than one curvature tensor.  Identical remarks apply to the region near $\sigma=\pi.$

The fact that our background has curvature singularities is perhaps a cause for concern.  For instance, one might worry that a local operator, $\Phi(x),$ in the $(2,0)$ theory might couple to curvature as
\begin{equation}
\delta L = \int d^{6}x \sqrt{g} ~R_{\mu\nu}R^{\mu\nu}~ \Phi(x)~. \label{worry}
\end{equation}
Near $\sigma=0, \pi$ where the singularity occurs, this yields an infinite source for the operator $\Phi(x)$ which could complicate our analysis.

Fortunately, we may rule out such couplings on general grounds.  Indeed, since the $(2,0)$ theory is scale invariant there is no dimensionful coupling constant that could appear in \eqref{worry} and hence the local operator $\Phi(x)$ necessarily has scaling dimension two.   Moreover it follows from the unitarity bounds of superconformal representation theory that such an operator is a free field \cite{Minwalla:1997ka}.  The interacting $(2,0)$ theory has no such free fields in its spectrum so \eqref{worry} cannot occur.  One may also rule out couplings of operators with spin to singular curvature tensors using an identical argument.

Instead of such singular couplings, we will ultimately argue that behavior of the theory near $\sigma=0,\pi$ can be modeled by a local Nahm pole boundary condition.

\subsection{Weyl Rescaling and Background Fields}
\label{rescale}

Having specified the metric field we can now proceed to fix the remaining backgrounds using the constraints of supersymmetry.  Instead of doing this directly it is instead more instructive to utilize the full power of the conformal invariance to simplify our analysis.

Conformal invariance of the $(2,0)$ theory means that the coupling to background supergravity is invariant (up to anomalies) under Weyl transformations.  Thus we may freely multiply the metric by an arbitrary function to simplify our problem.  A natural choice is to rescale \eqref{metric} by $f(\sigma)^{-2}.$  We obtain
\begin{equation}
ds^{2}  =   r^{2}\left[\frac{\ell^{2}}{4} \left(d\theta^{2}+\sin^{2}(\theta)d\phi^{2}\right) +\frac{1}{k^{2}}\left(d\psi+k\cos^{2}(\theta/2)d\phi\right)^{2}\right] + r^{2}\left[\frac{d\sigma^{2}+e^{A(x,y)}(dx^{2}+dy^{2})}{f(\sigma)^{2}}\right]. \label{wmetric}
\end{equation}
The metric \eqref{wmetric} is a now simply a product of a squashed Lens space $S^{3}_{\ell}/\mathbb{Z}_{k}$ with a three-manifold $M_{3}$ parameterized by $\sigma, x, y$.

Backgrounds of this type have been investigated in the context of the $3d$-$3d$ correspondence \cite{Cordova:2013cea, Lee:2013ida}.  In particular, it is known how to preserve supersymmetry on any metric of the form $S^{3}_{\ell}/\mathbb{Z}_{k} \times M_{3}$.  Therefore we may immediately use these results to specify the remaining supergravity background fields to ensure supersymmetry.

Let us present these background fields in the special case of trivial squashing $\ell \rightarrow 1$.  The case of general $\ell$ is technically more involved but conceptually similar and may be extracted from \cite{Cordova:2013cea}.  To specify the structure, we let $a$ denote a frame index on the three manifold $M_{3},$ and let $\hat{b}, \cdots$ denote R-symmetry indices valued in an $so(3)$ subgroup of the initial $so(5)$ symmetry.  This $so(3)$ subgroup contains the $so(2)$ which is used in topological twisting.\footnote{In particular, it is not the $so(3)$ R-symmetry of the class S $\mathcal{N}=2$ theory.}  The profiles of the remaining fields in Table \ref{table6dfields} are
\begin{equation}
T=0~, \hspace{.5in}D=\frac{2}{r^{2}}\delta_{\hat{a}\hat{b}}-\mathrm{Trace}~, \hspace{.5in} V_{a}=\frac{1}{2ir}\varepsilon_{a\hat{b}\hat{c}}-\frac{1}{2}\omega_{a \hat{b}\hat{c}}~.
\end{equation}

The most notable aspect of the above is the profile of the $R$-gauge field $V$.  The quantity $\omega_{a b c}$ is the spin connection on $M_{3}.$  The fact that the components of the $R$-gauge field are adjusted to compensate for the for the spin connection, means that the theory along $M_{3}$ is topologically twisted.  In particular, variations in the metric of $M_{3},$ which include both the Riemann surface metric, encoded by $A(x,y)$ and changes in the function $f(\sigma)$ are $Q$-exact.  Therefore changes in these quantities cannot alter answers to supersymmetric questions.  This feature persists to general values of the squashing parameter $\ell \neq 1.$  

We can also make contact with familiar features of the $S^{4}$ background of \cite{Pestun:2007rz}.  Let us denote a general supersymmetric spinor by $\eta$, and decompose it into representations of the local holonomy group $so(2)_{L}$ on the Riemann surface.  We encounter two pieces each of definite $2d$ chirality
\begin{equation}
\eta =\eta_{+}+\eta_{-}~.
\end{equation}
Since the $(2,0)$ is chiral, the $2d$ chirality is linked with the chirality of spinors in the remaining four-directions.  In our background on $S^{4}\times \Sigma$ we expect that the spinors become of pure chirality at the poles of $S^{4}$ where instantons or antiinstantons may be supersymmetrically localized.

In our setup, the poles of the $S^{4}$ are the loci $\sigma=0, \pi,$ and the thus we would like to understand how the supersymmetry parameters vary as a function of $\sigma$.  In the simple Weyl frame where the geometry looks like $S^{3}/\mathbb{Z}_{k}\times M_{3}$ the theory is twisted on $M_{3}$ and therefore the spinors behave as scalars, i.e. they are constant as functions of $\sigma, x, y$.  To recover the variation with respect to $\sigma$ we return to the orignial Weyl-frame of our calculation where the geometry contains a $S^{4}$.

Thus we Weyl rescale the metric by a factor $f(\sigma)^{2}$.  At the same time, we also perform a $\sigma$-dependent $R$-gauge transformation valued in the $so(2)$ subgroup of $so(5)$ that mixes with $so(2)_{L}$ to become the twisted $2d$ Lorentz group $so(2)_{\Delta}.$  If $\alpha(\sigma)$ denotes the exponentiated gauge transformation, then the result of the combined Weyl and $R$-gauge transformation is spinors $\eta_{\pm}$ whose $\sigma$ dependence is
\begin{equation}
\eta_{+}\sim \alpha(\sigma)\sqrt{f(\sigma)}~, \hspace{.5in}\eta_{-}\sim \alpha(\sigma)^{-1}\sqrt{f(\sigma)}~.
\end{equation}
Note that $\eta_{\pm}$ transform inversely with respect to the $R$-gauge transformation.  This is another consequence of twisting: the $2d$ chirality of the spinors under $so(2)_{L}$ is linked with their chirality under $so(2)_{R}$.  The desired properties of $\eta_{\pm}$ are now visible in a suitable choice of gauge.  For instance, the round $S^{4}$ discussed in \cite{Pestun:2007rz} corresponds to the choice $f(\sigma)=\sin(\sigma)$ and $\alpha(\sigma)=\sqrt{\cot(\sigma/2)}.$

In practice it is much more convenient to work in the Weyl frame where the squashed Lens space $S^{3}_{\ell}/\mathbb{Z}_{k}$ has constant size.  We thus proceed with dimensional reduction in this background.

\section{Complex Chern-Simons and Nahm Poles}

\subsection{Complex Chern-Simons Theory from (2,0) on $S^{3}_{\ell}/\mathbb{Z}_{k}$}

We have now arrived in a background of the form $S^{3}_{\ell}/\mathbb{Z}_{k} \times M_{3}.$  To understand the physics we first reduce on the smooth compact space  $S^{3}_{\ell}/\mathbb{Z}_{k}.$  This reduction has been studied in detail in \cite{Cordova:2013cea} (see also \cite{Yagi:2013fda, Lee:2013ida, Dimofte:2014zga} for closely related work).  The resulting effective field theory is complex Chern-Simons theory on $M_{3}$.

Complex Chern-Simons theory has been less studied than the more familiar Chern-Simons theory with compact gauge group.  Aspects of its geometric quantization were studied in \cite{Witten}, and investigations in perturbation theory were carried out in \cite{Dimofte:2009yn}.

To briefly recap of the results of  \cite{Cordova:2013cea}, the $(2,0)$ theory is labelled by an ADE Lie-algebra $\mathfrak{g}.$  After reduction on an $S^{3}_{\ell}/\mathbb{Z}_{k},$ the theory of zero modes consists of a $\mathfrak{g}_{\mathbb{C}}$ valued connection.
\begin{equation}
\mathcal{A}=A+iX~, \label{complexified}
\end{equation}
where $A,$ and $X$ are $\mathfrak{g}$ valued fields.

In the low-energy limit, the interactions among the degrees of freedom \eqref{complexified} are governed by the complex Chern-Simons Lagrangian
\begin{eqnarray}
S & = & \frac{q}{8\pi} \int \mathrm{Tr}\left({\cal A} \wedge d {\cal A} + \frac{2}{3} {\cal A} \wedge {\cal A} \wedge {\cal A}\right) + \frac{\tilde{q}}{8\pi} \int \mathrm{Tr}\left(\bar{\cal A} \wedge d\bar{\cal A} + \frac{2}{3} \bar{\cal A} \wedge \bar{\cal A} \wedge\bar{\cal A} \right) \label{SCS} \\
 & = & \frac{k}{4\pi} \int \mathrm{Tr}\left(A\wedge dA +\frac{2}{3} A^3 - X \wedge d_A X \right) + \frac{s}{2\pi} \int\mathrm{Tr} \left( \frac{1}{3} X^3- X \wedge F_A  \right)~, \nonumber
 \end{eqnarray}
 where $q = k + is$ and $\tilde{q} = k - is$ for $s$ not necessarily real. The level $k$ is quantized, while $s$ is a continuous variable set by the squashing as
 \begin{equation}
s=\sqrt{\ell^{2}-1}~.
\end{equation}
Note that both real and imaginary values of $s$ are accessible by appropriately tuning $\ell$.  In complex Chern-Simons both branches lead to unitary theories \cite{Witten}.

The most striking aspect of the above is that the dynamical variable \eqref{complexified} is subject to a complexified gauge redundancy.  Since this theory is obtained from a supersymmetric compactification, one would expect a supersymmetric theory with $\mathfrak{g}$ gauge redundancy, not a bosonic theory of $\mathfrak{g}_{\mathbb{C}}$ gauge fields.  As derived in detail in \cite{Cordova:2013cea}, the resolution of this seeming paradox is that the dimensional reduction produces in fact a gauge fixed form of the complex Chern-Simons theory.  The field $X$ is constrained to have vanishing $\mathfrak{g}$ covariant divergence
\begin{equation}
\nabla^{a}X_{a}=0~, \label{gaugefix}
\end{equation}
where $\nabla$ is the $A$ covariant derivative.  The condition \eqref{gaugefix} gauge fixes $\mathfrak{g}_{\mathbb{C}}\rightarrow \mathfrak{g},$ and correspondingly we also find Fadeev-Popov ghost fermions $\rho$.  Then, the gauge fixed action is indeed a supersymmetric $\mathfrak{g}$ gauge theory, and is equivalent to the claimed complex Chern-Simons theory.

This subtlety also serves to clarify the limitations of these results.  In fact, the fermion action for $\rho$ differs from the Fadeev-Popov action by a $Q$-exact term.  Therefore the relationship with complex Chern-Simons theory requires not just an appropriate low-energy limit to remove non-zero modes on $S^{3}_{\ell}/\mathbb{Z}_{k},$ but also a restriction to supersymmetric observables so that the $Q$-exact term may be ignored.  Thus, from now on we confine our attention to supersymmetric observables.  In addition to partition functions on compact manifolds, these supersymmetric observables include general Wilson lines in Chern-Simons theory which give rise to vertex operators in the Toda field theory.

\subsection{Boundary Conditions}
\label{sec:bc}

As a consequence of the previous section, we have reduced our problem to the study of Chern-Simons theory on the three-manifold $M_{3}$ which is a warped product of a Riemann surface $\Sigma$ and a line $\mathbb{R}.$   $M_{3}$ is equipped with the metric discussed in \S \ref{rescale}
\begin{equation}
ds^{2}= r^{2}\left[\frac{d\sigma^{2}+e^{A(x,y)}(dx^{2}+dy^{2})}{f(\sigma)^{2}}\right]~. \label{met2}
\end{equation}
Complex Chern-Simons theory is a topological field theory, and on a non-compact manifold such as $M_{3}$ the dynamics will reduce to a theory of edge modes on the boundary.  Our aim is to show that the theory of these edge modes is exactly the Toda field theory.

To reduce the theory to its boundary degrees of freedom it is necessary to precisely determine appropriate boundary conditions.  Let us therefore examine one boundary, say $\sigma=0$.  As $f(\sigma)$ vanishes linearly, the metric above becomes asymptotically hyperbolic, and we can anticipate a set of boundary conditions familiar from holography.  Indeed we will find boundary conditions relevant to perturbative gravity (viewed in Chern-Simons formalism) \cite{Coussaert:1995zp, Henneaux:2010xg}.

It is instructive to compare our problem to a closely related setup involving boundary conditions for D-branes in IIA string theory (appropriate for the case $\mathfrak{g}=\frak{su}(n)$), and their relation to M-theory geometries.  Our construction arises from a stack M5-branes wrapping the $S^{4}$ and we may reduce this to a configuration of D4-branes.  This reduction is useful because we may understand the origin of the Chern-Simons fields in this language.  Specifically, $\mathcal{A}=A+iX$ is composed of the worldvolume gauge field and a triplet of twisted scalars of the D4-brane theory.

In order to preserve supersymmetry, our reduction from the $(2,0)$ theory to 5d Super-Yang Mills must be carried out in a particular way, by reduction on the Hopf fiber of $S^{3}_{\ell}/\mathbb{Z}_{k}.$  In the case the round background this may be understood simply in terms of supergroups.  The round $S^{4}$ background preserves $osp(2|4)$ supersymmetry, where $sp(4)\cong so(5)$ is the isometry group of the $S^{4}$.  To carry out a supersymmetric reduction along a circle we must find a $u(1)$ subgroup of the $so(5)$ isometry which acts trivially on some odd generators.  We accomplish this by picking an $su(2|1)\times su(2)\subset osp(2|4)$ and reducing with respect to the $u(1)\subset su(2).$  This $u(1)$ rotates the Hopf fiber, while the remaining $su(2)$, under which the spinors are charged, rotates the base $S^{2}$ of the fibration.

Since the reduction to 5d Yang-Mills preserves supersymmetry, we can reliably use this description to determine the boundary conditions.  Near $\sigma=0$ the Hopf fiber, which plays the role of the M-theory circle, collapses.  Moreover this locus is codimension four in the ambient M-theory geometry, i.e. it is a point, the pole of the $S^{4}$.  Therefore in type IIA language the analogous boundary condition is a that of D6 branes which meet the D4-brane stack at $\sigma=0$.

The boundary conditions which describe D4's ending on D6's have been investigated in detail in \cite{Gaiotto:2008sa, Gaiotto:2008ak }.  They are Nahm pole boundary conditions.  They are characterized by a divergence in the triplet of twisted scalars valued in an $su(2)$ subalgebra of $\mathfrak{g}$.  Explicitly, given a triplet $T_{i}$ of generators of $\mathfrak{g}$ obeying $[T_{i}, T_{j}]=\epsilon_{ijk}T_{k}$ the Nahm pole boundary condtition in flat space is that as $\sigma \rightarrow 0,$
\begin{equation}
X_{i}\rightarrow\frac{1}{\sigma}T_{i}~, \label{nahm}
\end{equation}
In particular, \eqref{nahm} implies
\begin{equation}
 \frac{dX_{i}}{d\sigma}+\epsilon_{ijk}X_{j}X_{k}=0~.
\end{equation}

Nahm pole boundary conditions are thus labelled by a conjugacy class of homomorphism from $\frak{su}(2)$ to $\mathfrak{g}$ specifying the triplet $T_{i}$ up to gauge redundancy.  These homomorphisms are in turn fixed once the image of the raising operator $T_{+}=T_{1}+iT_{2}\in su(2)_{\mathbb{C}}$ is known. For the case of case of $\mathfrak{g}=su(n)$ the image of $T_{+}$ is an $n\times n$ nilpotent matrix which may be put into Jordan canonical form.  The Jordan canonical form results in a partition of $n$ into $p$ parts $n_{i}.$  The number $p$ is then interpreted as the number of D6-branes, and the $n_{i}$ specify how many D4-branes end on the $i$-th D6-brane.

In our case the M-theory geometry is $S^{4}/\mathbb{Z}_{k}$.  In particular when $k=1$ this geometry is smooth so there is exactly one D6-brane at our boundary. Therefore we choose the image of $T_{+}$ to be a maximal length Jordan block.  With $\mu_{i}$ some non-zero real constants we have
\begin{equation}
T_{+}\rightarrow  \left(\begin{array}{ccccc} 0 &\mu_{1}  & 0  &\cdots & 0\\
 0 &0&  \mu_{2}&  \cdots & 0 \\
 \vdots & \vdots  & \vdots & \ddots & \vdots \\
 0 & 0 & 0 & \cdots & \mu_{n-1}\\
 0 & 0 & 0 &  \cdots & 0 \end{array}\right), \label{nilpexp}
\end{equation}
For later convenience, we may choose the normalizations of the matrix elements $\mu_{i}$ above to match with standard eigenvalue conventions for the $su(2)$ Cartan Generator. If we define a lowering operator as $T_{-}=T^{\dagger}_{+},$ and a Cartan generator $H$ via $[T_{+},T_{-}]=2H,$ then we fix conventions so that the eigenvalues of $H$ are half-integral.  For the case of $k\neq1$ we continue to use this type of Nahm pole boundary condition characterized by a maximal Jordan block.

For the general case of the $\mathfrak{g}$ an arbitrary ADE Lie algebra, we proceed as follows.  According to the Jacob-Morozov theorem, conjugacy classes of homomorphisms from $su(2)$ to $\mathfrak{g}$ are determined by the decomposition of the fundamental representation of $\mathfrak{g}$ into representations of $su(2)$, and moreover every dimensionally consistent decomposition is associated to some conjugacy class of homomorphism.  In the case $\mathfrak{g}=su(n)$ this is equivalent to the partition of $n$ specified above.  We then choose our Nahm-Pole boundary condition to be such that the fundamental representation of $\mathfrak{g}$ is a single irreducible representation of $\frak{su}(2)$.

The above analysis of Nahm pole boundary conditions is directly relevant for $D4$-branes in flat space with an explicit M-theory embedding of the full supergravity background.  For our problem we must now determine an analogous boundary condition which applies to the full complex Chern-Simons theory.  It is easy to surmise the correct behavior from two simple requirements.  First the complex connection $\mathcal{A}$ must asymptotically have a linear divergence characterized by the Nahm pole, and second the connection must be flat.  The solution to these conditions is that for $\sigma\rightarrow0$
\begin{equation}
\mathcal{A}\rightarrow\frac{d\sigma}{\sigma}H+\frac{du}{\sigma}T_{+}~. \label{backgroundA}
\end{equation}

In \eqref{backgroundA} $u$ is a holomorphic boundary coordinate on $\Sigma$.  It is convenient to view this variable as a Euclidean continuation of a light-cone coordinate in Minkowski signature.  This perspective is useful for determining the correct boundary behavior of $\bar{\mathcal{A}}$.  Indeed, in Lorentzian signature $u$ is a real variable, so we demand that
\begin{equation}
\mathcal{\bar{A}}\rightarrow\frac{d\sigma}{\sigma}H+\frac{du}{\sigma}T_{+}~. \label{backgroundAbar}
\end{equation}

It is instructive to compare the above boundary behavior of the connection to more standard boundary conditions that are frequently imposed in Chern-Simons theory \cite{Elitzur:1989nr}.    In general, a simple criterion is to impose that the boundary terms in the equation of motion vanish.  Modulo terms which vanish by bulk equations of motion, the variation of the Chern-Simons action is
\begin{equation}
\delta S \sim \int _{\partial M_{3}}(\delta \mathcal{A} \wedge \mathcal{A})~.
\end{equation}
Thus, one possible boundary condition is to choose a component of $\mathcal{A}$ along the boundary to vanish.  In our case the boundary coordinates are $u$ and $\bar{u}$ and we have
\begin{equation}
\mathcal{A}_{\bar{u}}=\bar{\mathcal{{A}}}_{\bar{u}}=0~. \label{ordinarybc}
\end{equation}
The boundary conditions \eqref{backgroundA}-\eqref{backgroundAbar} are a strengthening of \eqref{ordinarybc} which further constrains the remaining components of the connection.  This perspective will be useful in the following.  It would be interesting to reformulate our analysis directly using geometric quantization as in \cite{Verlinde:1989ua}.

Finally, let us also remark on the boundary conditions on the other side of the interval at $\sigma=\pi$.  Since the total space of the $S^{4}_{\ell}/\mathbb{Z}_{k}$ is compact without net D6 brane charge the boundary condition at $\sigma=\pi$ must be described by a Nahm pole for anti-D6-branes.  This means that the connection obeys constraints identical to \eqref{backgroundA}-\eqref{backgroundAbar}, except the chirality is reversed, $u\leftrightarrow \bar{u},$ and the positive and negative roots of $\frak{g}$ are exchanged, $T_{+}\leftrightarrow T_{-}$.

\subsection{Fluctuation Modes}
\label{sec:fluc}

We now describe fluctuating edge modes in the connection $\mathcal{A}$ which obey the boundary conditions \eqref{backgroundA}-\eqref{backgroundAbar}.  We again focus on a single boundary $\sigma=0$ and will find a chiral edge mode spectrum.  The additional boundary at $\sigma=\pi$ gives rise to edge modes of opposite chirality.

The boundary condition \eqref{backgroundA} involves a flat gauge field $\mathcal{A}$.  Therefore locally, we may write $\mathcal{A}=G^{-1}dG$ for a $G_{\mathbb{C}}$-valued matrix $G$.  Using the commutation relation
\begin{equation}
[T_{+},H^{n}]= \left(\sum_{k=1}^{n}\binom{n}{k}(-1)^{k}H^{n-k}\right)T_{+}~,
\end{equation}
one finds that
\begin{equation}
G=\exp\left(uT_{+}\right)\exp\left(\phantom{\frac{u}{w}}\hspace{-.18in}\log(\sigma)H\right)~.
\end{equation}

To describe fluctuating fields we relax our constraint on $G$ to take the more general form
\begin{equation}
G=g(u)\exp\left(\phantom{\frac{u}{w}}\hspace{-.18in}\log(\sigma)H\right)~.
\end{equation}
We wish to understand the constraints on $g(u)$ so that the above is compatible with the boundary conditions.  Before doing so however, we must explain why it is appropriate to restrict $g(u)$ to be independent of $\bar{u}$.

The restriction to holomorphic $g$ can be justified by considering the constraints of Chern-Simons theory on a manifold with boundary \cite{Witten:1988hf, Elitzur:1989nr}.  The variable $\mathcal{A}_{\bar{u}}$ may be viewed as a Lagrange multiplier, and the path integral over $\mathcal{A}_{\bar{u}}$ imposes that, for each value of $\bar{u}$, the connection on the $(u,\sigma)$ space is flat.  Therefore we may express $\mathcal{A}$ as
\begin{equation}
\mathcal{A}=G^{-1}\tilde{d}G~, \label{aflat}
\end{equation}
where $G$ depends on $u, \bar{u}$ and $\sigma$ and $\tilde{d}$ is the exterior derivative on the $(u,\sigma)$ space.  This ansatz solves the constraints but it is redundant.  Multiplying $G$ by a matrix valued function of $\bar{u}$ leaves the dynamical variable $\mathcal{A}$ invariant.  Therefore to describe physical boundary modes in the connection $\mathcal{A}$ we restrict our attention to holomorphic $g$.

Our next task is to understand how the boundary degrees of freedom contained in $g(u)$ are constrained by the Nahm pole boundary condition \eqref{backgroundA}.  This analysis has been carried out in \cite{Gaiotto:2008sa}.  We introduce a Cartan decomposition of the algebra into positive roots, $\Delta_{+},$ negative roots, $\Delta_{-},$ and a Cartan subalgebra $\Delta_{0}.$  We then write
\begin{equation}
g^{-1}dg=\left(\sum_{i\in \Delta_{+}}\chi_{i}^{+}R_{i}^{+}+\sum_{i\in \Delta_{-}}\chi_{i}^{-}R_{i}^{-}+\sum_{i\in \Delta_{0}}\chi_{i}^{0}R_{i}^{0}\right)du~,
\end{equation}
where the $\chi$'s depend only on the variable $u$ and $R$'s denote the generators of the Lie algebra in the indicated subspace.  The connection $\mathcal{A}$ derived from $G$ is
\begin{equation}
\mathcal{A}=\frac{d\sigma}{\sigma}H+\exp(-\log(\sigma)H)(g^{-1}dg)\exp(\log(\sigma)H). \label{inter1}
\end{equation}
To simplify \eqref{inter1} note that  the various generators $R_{i}$ obey simple commutation relations with the element $H$
\begin{equation}
[H,R_{i}]=h_{i}R_{i}, \label{commrels}
\end{equation}
where the quantities $h_{i}$ are positive negative or vanishing depending on whether $R_{i}$ is a member of $\Delta_{+}, \Delta_{-},$ or $\Delta_{0}$ respectively.  Therefore, the connection may be written as
\begin{equation}
\mathcal{A}=\frac{d\sigma}{\sigma}H +\sum_{i}\frac{\chi_{i}(u) R_{i}}{\sigma^{h_{i}}}du~.
\end{equation}

Now we impose the boundary condition, namely we demand that the singular part of $\mathcal{A}$ is given exactly by the Nahm pole boundary condition \eqref{backgroundA}.  Whence, the terms in the sum above where $h_{i}>0$ must vanish except for the single term where $R_{i}=T_{+}.$  Hence we have
\begin{equation}
\mathcal{A}=\frac{d\sigma}{\sigma}H +\frac{du}{\sigma}T_{+}+\sum_{i\in \Delta_{0}\cup \Delta_{-}}\frac{\chi_{i}(u) R_{i}}{\sigma^{h_{i}}}du~.
\end{equation}
We may further simplify the form of the connection by using gauge transformations.  Consider an infinitesimal gauge transformation with gauge parameter $\lambda$ of the form
\begin{equation}
\lambda=\sum_{i \in \Delta_{-}}\frac{\lambda_{i}(u)R_{i}}{\sigma^{h_{i}}}~.
\end{equation}
Such a gauge transformation preserves the $d\sigma$ component of $\mathcal{A}$ and vanishes at $\sigma \rightarrow 0,$ hence it implies that various fluctuation modes $\chi_{i}$ are in fact redundant.  To deduce which modes may be removed, view the Lie algebra $\mathfrak{g}$ as a representation of the $sl(2,\mathbb{C})$ subalgebra generated by $T_{+}, H, T_{-}.$  In general, the representation in question is reducible, for instance with $\mathfrak{g}=sl(n,\mathbb{C})$ this representation splits as a direct sum
\begin{equation}
sl(n,\mathbb{C})|_{sl(2,\mathbb{C})}\equiv \mathbf{3} \oplus \mathbf{5} \oplus \mathbf{7} \oplus \cdots \oplus \mathbf{2n-1}.
\end{equation}
A simple way to verify this decomposition is to identify the lowest weight vectors in each irreducible summand.  These are precisely those vectors which commute with the element $T_{-}$ and for the Nahm boundary condition defined by \eqref{nilpexp} they are precisely the powers $T_{-}^{\ell}$ where $\ell$ ranges from $1$ to $n-1$.

Now examine a gauge transformation generated by a vector of definite weight in the decomposition above.  The commutator with the raising generator $T_{+}$ appearing in $\mathcal{A}$ shifts the weight up by one and moreover, because of the Nahm pole, also decreases the power of $\sigma$ down by one.  Thus using such a gauge transformation, we may eliminate those $\chi$ modes in $\mathcal{A}$ proportional to a weight vector which is not lowest.  What remains then is a sum over lowest weight vectors.

Again returning to the case of $sl(n,\mathbb{C})$ the expansion of the connection can be made very explicit using our understanding of the lowest weight vectors as powers of $T_{-}.$  We have
\begin{equation}
\mathcal{A}=\frac{d\sigma}{\sigma}H +\frac{du}{\sigma}T_{+}+\sum_{j=1}^{n-1}\chi_{j}(u)(T_{-})^{j}\sigma^{j} du~.
\end{equation}
There are thus $n-1$ fluctuating chiral fields supported at the boundary.  In the case of general Lie algebra $\mathfrak{g}$ we find rank($\mathfrak{g}$) chiral edge modes on the boundary.  In the next section we will argue that these are exactly the chiral Toda fields.

\section{The Constrained WZW Model and Toda}

\subsection{The Complex WZW Model}

Having understood the boundary conditions and fluctuating fields in the language of Chern-Simons theory, we must now understand the effective action for these modes.     The boundary conditions discussed in section \ref{sec:bc} include in particular
\begin{equation}
\mathcal{A}_{\bar{u}}=0~, \hspace{.5in} \mathcal{\bar{A}}_{\bar{u}}=0~.\label{bc1new}
\end{equation}

Boundary conditions of this type reduce the Chern-Simons theory to an edge mode WZW model.  Let us briefly review this construction following \cite{Witten:1988hf, Elitzur:1989nr}.  As in \eqref{aflat}, we view the Chern-Simons path-integral over the $\bar{u}$ components of the connection as a Lagrange multiplier enforcing flatness of the connection on the $(u,\sigma)$ space.  We therefore change variables to a group valued field
\begin{equation}
\mathcal{A}=G^{-1}\tilde{d}G~,
\end{equation}
where as before $\tilde{d}$ is the exterior derivative on $(u,\sigma)$.  The Chern-Simons action may then be directly integrated to produce an action for $G$.  This is the chiral WZW action
\begin{equation}
S =n I_{WZW} = \frac{n}{4 \pi} \int_{\partial M_{3}} \textrm{Tr}\left(G^{-1}\partial_{u} G G^{-1} \partial_{\bar{u}}G \right) + \frac{n}{12\pi} \int_{M_{3}} \textrm{Tr}\left(G^{-1} dG \wedge G^{-1} dG \wedge G^{-1} dG \right)~.
\end{equation}
Since the value of $G$ in the interior may be changed by gauge transformation the action depends only on the boundary values of $G$.  The theory on a single boundary is chiral for the reasons discussed in section \ref{sec:fluc}. More specifically, we may consider transforming $G$ as
\begin{equation}
G\rightarrow V(u)G\tilde{V}(\bar{u})~.
\end{equation}
If we view the $\bar{u}$ direction as time, then the transformations by $\tilde{V}$ are gauge redundancies which allow us to eliminate the $\bar{u}$ dependence.  Meanwhile, the transformations by $V$ are global symmetries.

The analysis above applies both for standard compact gauge groups and for the less standard complex Chern-Simons theory relevant here.  In the former case the parameter $n$ is quantized and equal to the bulk Chern-Simons level.  In the complex case instead the chiral WZW action is
\begin{equation}
S=\frac{q}{2} I_{WZW}(G)+\frac{\tilde{q}}{2}I_{WZW}(\bar{G})~,
\end{equation}
where $q$ and $\tilde{q}$ are the complex levels of the Chern-Simons theory introduced in \eqref{SCS}.  Note in particular that as a consequence of our boundary conditions, both $G$ and $\bar{G}$ have physical degrees of freedom that depend only on $u$.

The arguments given so far are correct for the case of a manifold $M_{3}$ with a single boundary component.  However, our construction which arises from the $(2,0)$ theory on a $S^{4}$ involves an $M_{3}$ with two identical boundary components and it is simple to generalize to this situation.    If the previous considerations are taken to apply to the boundary Riemann surface localized at $\sigma=0,$ then the boundary at $\sigma=\pi$ gives rise to another current algebra of opposite chirality \cite{Elitzur:1989nr}.  Moreover, Wilson lines in the Chern-Simons theory stretched between the two boundary components produce local vertex operators.  The result of the Chern-Simons theory on the manifold with two-boundaries, in the presence of the boundary condition \eqref{bc1new} is therefore the full non-chiral $\mathfrak{g}_{\mathbb{C}}$ WZW model.

\subsection{Current Constraints and Complex Toda }
\label{sec:complextoda}

We have arrived at a complexfied model of currents, however this is not the end of our analysis.  The reason is that the Nahm pole boundary conditions of section \ref{sec:bc} are stronger than those used to reach the WZW model.  We will now argue that these additional restrictions imposed by the boundary conditions can be phrased as constrains on the currents.

Since the currents are chiral, to see the constrains it suffices to focus on a single boundary.  We then proceed as before and express the connection as $\mathcal{A}=G^{-1}\tilde{d}G$ with
\begin{equation}
G=g(u)\exp(\log(\sigma)H)~.
\end{equation}
Using a gauge transformation, we may remove the $\sigma$ dependence from the matrix above.  We are left with the dynamical boundary variable $g(u)$ and aim to describe the constraints of the Nahm pole in terms of the holomorphic current
\begin{equation}
J(u)=g^{-1}dg~.
\end{equation}
This is precisely the analysis we have carried out in section \ref{sec:fluc}.

The projection of the currents onto the positive roots $\Delta_{+}$ of the Lie algebra are restricted to produce the Nahm pole proportional to $T_{+}.$  This generator can be expressed as a sum of the positive simple roots with non-zero coefficients.  Therefore, letting $J(R_{i}^{+})$ denote the projection of the current onto the $i$-th positive root, we have
\begin{equation}
J(R_{i}^{+})=\mu_{i}~,
\end{equation}
where $\mu_{i}$ is a constant which is non-zero for every simple positive root, but vanishes otherwise.  We must additionally take into account the restrictions on the currents imposed by bulk gauge redundancy in the presence of the Nahm pole.  These enable us to set to zero the currents along Cartan directions $R_{i}^{0}$.  Therefore
\begin{equation}
J(R_{i}^{0})=0~.
\end{equation}

By taking into account the other boundary we find analogous constraints on the anti-chiral currents $\tilde{J}=\partial_{\bar{u}}gg^{-1}$, with the roles of positive and negative roots reversed
\begin{equation}
\tilde{J}(R_{i}^{-})=\nu_{i}~, \hspace{,5in}\tilde{J}(R_{i}^{0})=0~,
\end{equation}
where $\nu_{i}$ are non-zero exactly when the root is simple.

Current constraints of this type have been studied in detail for the $\mathfrak{sl}(n,\mathbb{R})$ WZW model \cite{Forgacs:1989ac, Balog:1990mu}.  In that case it is known that the constraints reduce the WZW model to the Toda field theory.  The same setup also arises in studies of pertrubative three-dimensional gravity on AdS space with Brown-Henneaux boundary conditions \cite{Coussaert:1995zp, Henneaux:2010xg}.  Gravity in 2+1 dimensions is topological and is perturbatively equivalent to $\mathfrak{sl}(2,\mathbb{R}) \times \mathfrak{sl}(2,\mathbb{R})$ Chern-Simons theory.  Moreover the natural holographic boundary conditions are exactly of the Nahm pole type.  This gives rise holographically to Liouville theory on the boundary of spacetime where each factor of the gauge group produces a chiral half of the Liouville system.

Let us review the relevant aspects of these results following \cite{Forgacs:1989ac}.  We first consider the rank one situation, related to $\frak{sl}(2)$ algebras, and then subsequently generalize.

It is useful to express the field $g$ via a Gauss decomposition.  Thus we write
\begin{equation}
g=\left(\begin{array}{cc} 1 & X\\ 0 & 1 \end{array}\right) \left(\begin{array}{cc} e^{\Phi/2} & 0\\ 0 & e^{-\Phi/2} \end{array} \right) \left( \begin{array}{cc} 1 & 0 \\ Y & 1 \end{array} \right)~.
\end{equation}
In the case of $\frak{sl}(2,\mathbb{R})$ all fields are real, while in the $\frak{sl}(2,\mathbb{C})$ case relevant to our analysis all fields are complex.   We now evaluate the currents and enforce the constraints
\begin{equation}
J_+ = e^{-\Phi} \partial X = \mu~,  \hspace{.5in}J_0 = \frac{1}{2} \partial \Phi + Y e^{-\Phi} \partial X=0 ~,
\end{equation}
\begin{equation}
\tilde{J}_- = e^{-\Phi} \bar{\partial} Y = \nu~,  \hspace{.5in}\tilde{J}_{0}=\frac{1}{2} \bar\partial \Phi + X e^{-\Phi} \bar\partial Y=0 ~. \nonumber
\end{equation}
The equations for $J_+$ and $\tilde{J}_-$  imply that $X$ and $Y$ are not independent fields, and only $\Phi$ remains. The vanishing of $J_0$ and $\tilde{J_0}$, together with current conservation, then result in the Liouville equation of motion
\begin{equation}
 \partial\bar\partial \Phi+2\mu \nu e^\Phi =0~.
\end{equation}
Alternatively, one may solve the constraints without using current conservation and obtain the Liouville action directly from the WZW action.  Note that the values of $\mu$ and $\nu$ are irrelevant as long as they are both positive since they may be absorbed by a shift of $\Phi$.

An important aspect of this construction is that we can understand the unusual conformal transformation of the Liouville field.  The current constraints involve parameters $\mu$ and $\nu$ which are dimensionful, and thus break the original scale symmetry generated by left and right moving Virasoro generators $L_{0}$ and $\bar{L}_{0}$.  To restore conformal invariance, we modify the scale transformations to
\begin{equation}
L_{0}'=L_{0}-H~,\hspace{.5in}\bar{L}_{0}'=\bar{L}_{0}+\bar{H}~, \label{modifiedl0}
\end{equation}
with $H, \bar{H}$ the Cartan generators discussed in section \ref{sec:bc}.  The modified generators then assign the currents which are fixed to constants to have scaling dimension zero, and can be extended to a full Virarsoro symmetry of the constrained system \cite{Forgacs:1989ac}.  The cost of this redefinition is that the field $\Phi$ no longer transforms homogeneously.  Under transformation by the Cartan elements, and hence with the modified scale transformations, $\Phi$ shifts.

We can now easily generalize to other Lie algebras.  Again we factorize $g$ into a pieces with positive roots, negative roots, and Cartan elements
\begin{equation}
g=\exp(X_{i}R_{i}^{+})\exp\left(\Phi_{i}R_{i}^{0}\right)\exp(Y_{i}R_{i}^{-})~.
\end{equation}
As before, in the $\frak{g}_{\mathbb{C}}$ case relevant to us, all fields are complex.  The constraints for the positive and negative roots may be expressed as
\begin{eqnarray}
\exp(-X_{i}R_{i}^{+})\partial(X_{i}R_{i}^{+})& = &\sum_{i \in \Delta_{+}}\mu_{i}R_{i}^{+}\exp\left(C_{ij}\Phi_{j}\right)~,\\
\bar{\partial}(Y_{i}R_{i}^{-})\exp(-Y_{i}R_{i}^{-}))& = &\sum_{i \in \Delta_{-}}\nu_{i}R_{i}^{-}\exp\left(C_{ij}\Phi_{j}\right)~.
\end{eqnarray}
Therefore, as in the $\frak{sl}(2)$ case, the constraints enable us to eliminate the fields $X_{i}$ and $Y_{j}$ in terms of $\Phi_{i}$.  The vanishing of the Cartan direction of the currents then yields the Toda equation of motion
\begin{equation}
\partial \bar{\partial}\Phi_{i}+2\mu_{i}\nu_{i}\exp\left(C_{ij}\Phi_{j}\right)=0~,
\end{equation}
exactly as derived from the action \eqref{complexToda}.

This non-chiral complex Toda theory is our final result for the effective theory of the $\frak{g}$-type (2,0) theory reduced on $S^{4}_{\ell}/\mathbb{Z}_{k}.$

\section{Duality of Complex Toda and ParaToda + Coset}
\label{sec:duality}

After the analysis of the previous section we have reduced the edge modes of complex Chern-Simons theory to complex Toda field theory.  In this section we propose that this complex Toda theory is dual to Para-Toda theory plus a decoupled coset model.  This duality relates our results to the AGT conjecture \cite{Alday:2009aq, Wyllard:2009hg} and its orbifold generalization \cite{Nishioka:2011jk}.
The conjectured duality pertains to the case $\frak{g}=\frak{su}(n).$  It states the equivalence
\begin{equation}
\mathrm{complex~ Toda}(n,k,s)\leftrightarrow \mathrm{real~paraToda}(n,k,b)+\frac{\hat{\frak{su}}(k)_{n}}{\hat{\frak{u}}(1)^{k-1}} ~,\label{duality2}
\end{equation}
where the two models on the right-hand-side of the above are decoupled, and the map between parameters is
\begin{equation}
b=\sqrt{\frac{k-is}{k+is}}~. \label{paramrel}
\end{equation}
The actions for each model are given in \eqref{complexToda} and \eqref{paratoda}.

As a first step towards motivating this duality let us begin by explaining why complex Toda in fact describes two decoupled CFTs.  This can easily be understood from the chiral sector of the theory.  The left-moving algebra of holomorphic constrained currents is inherited from the complex WZW model.  In the WZW model the currents have scaling dimension one, however in the Toda theory the dilatation generator is modified to $L_{0}'=L_{0}-H$ as in \eqref{modifiedl0}.  In particular the current projected onto the Lie algebra direction $T_{-}$ therefore has (left) scaling dimension two and hence is a holomorphic energy-momentum tensor.  However, because the current in the WZW model is complexified this argument in fact produces two independent spin two currents.  Therefore on general grounds we expect two decoupled field theories.

At a heuristic level, one reason to anticipate a duality such as \eqref{duality2} is due to bose-fermi equivalence in two-dimensional field theory.  As explained in section \ref{sec:complextoda}, the relationship between the WZW model and Toda involves an exponential change of variables.  Therefore the fields $\Im(\Phi_{i})$ are naturally periodic bosons and may be dualized.  The claim of \eqref{duality2} is that the result of this procedure is the paraToda theory plus the parafermions describing the decoupled coset model.

This intuition may also be used to derive the proposed map \eqref{paramrel} between the complex Toda parameters and those of the paraToda system.  Consider for simplicity the case of complex Liouville theory where the action is
\begin{equation}
S=\frac{k+is}{8\pi}\int\left( \partial \Phi \bar{\partial} \Phi+e^{\Phi}\right)+\frac{k-is}{8\pi}\int\left( \partial \bar{\Phi} \bar{\partial} \bar{\Phi}+e^{\bar{\Phi}}\right)~.\label{clvl}
\end{equation}
In the field variables $\Re(\Phi)$ and $\Im(\Phi)$ the kinetic terms are non-canonical.  We can make them standard by changing basis to
\begin{equation}
\phi=\sqrt{\frac{k^{2}+s^{2}}{k}}\Re(\Phi)~, \hspace{.5in}\theta=\sqrt{k}\Im(\Phi)+\frac{s}{\sqrt{k}}\Re(\Phi)~.\label{canonicalvars}
\end{equation}
Note in particular that the variable $\theta$ is periodic with $\theta \cong \theta +2\pi \sqrt{k}$.  In these variables, the action \eqref{clvl} can be recast as
\begin{equation}
\frac{1}{4\pi}\int \left(\partial \phi \bar{\partial}\phi-\partial\theta\bar{\partial}\theta\right) +\frac{k+is}{8\pi}\int e^{i\theta/\sqrt{k}}e^{b \phi/\sqrt{k}}+\frac{k-is}{8\pi}\int e^{-i\theta/\sqrt{k}}e^{b^{-1} \phi/\sqrt{k}}~, \label{paraLiouville}
\end{equation}
where $b$ is given by \eqref{paramrel}.

The classical manipulations used to arrive at the action \eqref{paraLiouville} are valid in the weak-coupling regime where either $b$ or $b^{-1}$ is small.   In that case we view the above as a semiclassical version of the paraLiouville system where $\phi$ is the Liouville field and we fermionize $\theta$ into the parafermions and decoupled coset.

Note also that with this identification of parameters, the complex Toda theory is naturally related to an analytic continuation of paraToda theory at complex $b$.  (Such analytic continuations have been studied for instance in \cite{Harlow:2011ny}.)  More standard real values of $b$ may be achieved by tuning $s$ to be purely imaginary.  Recalling the relationship between $s$ and the squashing parameter $\ell$ as $s=\sqrt{\ell^{2}-1}$ we see that both real and complex $b$ naturally occur in compactification of the $(2,0)$ theory on $S^{4}_{\ell}/\mathbb{Z}_{k}$.

These remarks also clarify the meaning of the wrong sign kinetic term in \eqref{paraLiouville} for the $\theta$ field.  In the case dual to unitary paraToda plus coset, $s$ is pure imaginary and the field variable $\theta$ is on an unusual contour of integration dictated by \eqref{canonicalvars} and the reality of both $\Re(\Phi)$ and $\Im(\Phi)$.  More general values of $s$ should be interpreted as analytic continuations of this situation and treated similarly to analytic continuations of Chern-Simons theory \cite{Witten:2010cx}.

Finally we can get another motivation for the parameter identification \eqref{paramrel} by considering the central charge.  The central charge of complex Chern-Simons edge modes has been computed in \cite{Witten}.  This analysis is not directly relevant to our construction because our edge mode theory results from constraints which break the original conformal symmetry of the boundary WZW theory.  In other words the scaling generators are shifted as in \eqref{modifiedl0}.  Moreover, since the constraints gauge some of the currents, we must couple to an appropriate ghost system to incorporate the gaugeing at the quantum level as in \cite{Feher:1992yx}.

Because of these issues, a proper calculation of the central charge of our system is involved.  Nevertheless we can provide one simple consistency check.  Indeed, both the shift and quantum ghost system do not involve the coupling $b$.  Therefore for large or small $b$ these complications may be ignored and we expect a match between the central charge of the complex Chern-Simons edge mode theory and the paraToda plus coset.  For the former, we have from \cite{Witten}\footnote{Reference \cite{Witten} studied a boundary condition where the edge modes are non-chiral.  For our application we have utilized a distinct boundary condition with chiral edge modes.  The central charge quoted in \eqref{cscentral} is then $c_{L}-c_{R}$ of \cite{Witten}.}
\begin{equation}
c_{WZW_{\mathbb{C}}}=2h^{\vee}d_{\frak{g}}\left(\frac{1}{q}+\frac{1}{\tilde{q}}\right)=\frac{4h^{\vee}d_{\frak{g}}}{k}\left(\frac{k^{2}}{k^{2}+s^{2}}\right)~, \label{cscentral}
\end{equation}
where $h^{\vee}$ is the dual coxeter number and $d_{\frak{g}}$ is the dimension of the (real) Lie algebra.\footnote{  For $\frak{su}(n)$ these are $h^{\vee}=n$ and $d=n^{2}-1$.}  By contrast the central charge of the ParaToda plus coset theory is \cite{Nishioka:2011jk}
\begin{equation}
c_{paraToda}+c_{coset}=\frac{n^{3}-n}{k}(b+b^{-1})^{2}+k(n-1)~. \label{cparacoset}
\end{equation}
The equations \eqref{cscentral} and \eqref{cparacoset} agree for large or small $b,$ with $b$ given in terms of $k$ and $s$ as in \eqref{paramrel}.  Similar observations have been made in \cite{Vafa:2015euh} to argue for a relationship between complex Chern-Simons and Toda.

One interesting aspect of the complex Liouville theory (or more generally the complex Toda theory) is its approximate factorization into degrees of freedom described by $\Phi$ and $\bar{\Phi}$ respectively.  Although the action \eqref{clvl} is naturally a sum, these variables are not decoupled.  Instead they interact through the fact that their contours of integration in field space are linked.

To elaborate on these ideas, from the complex Toda action it is natural to define couplings $b_{+}$ and $b_{-}$ which are the parameters for two Toda theories described by $\Phi_{i}$ and $\bar{\Phi}_{i}$
\begin{equation}
b_{+}^{2}=\frac{2}{k+is}~, \hspace{.5in}b_{-}^{2}=\frac{2}{k-is}~,
\end{equation}
If these models were truly decoupled one would expect the central charges to add leading to 
\begin{eqnarray}
c_{naive} & = &(n^{3}-n)(b_{+}+b_{+}^{-1})^{2}+(n^{3}-n)(b_{-}+b_{-}^{-1})^{2}+ 2(n-1) \nonumber\\
& = & \frac{4(n^{3}-n)}{k}\left(\frac{k^{2}}{k^{2}+s^{2}}\right)+(4+k)(n^{3}-n)+2(n-1)~.
\end{eqnarray} 
The difference between $c_{naive}$ and the exact central charge \eqref{cparacoset} is a $k$ and $n$ dependent shift.  This is consistent with the fact that the models are coupled through their contour of integration in field space since the difference in central charges does not depend on the continuous coupling $s$.

These manipulations also show that the proposed duality is in a sense a strong-weak duality.  The naive limit of weak coupling of the complex Toda theory is when both $b_{+}$ and $b_{-}$ tend to zero with fixed $k$.  However in this limit the $b$ parameter of the paraToda theory which is $b_{+}/b_{-}$ tends to $i$.  Conversely, when $b$ or $b^{-1}$ tends to zero, one of the Toda systems is strongly interacting.

\subsection{Additional Evidence at Small $k$}
To get more specific evidence for the duality, we now consider the case of small $k$.  We focus on the rank one case of $\frak{sl}(2)$ theory where the duality concerns Liouville theory.  The higher rank version which pertains to Toda is a natural generalization of these ideas.

\subsubsection{The Case $k=1$}

The original AGT conjecture \cite{Alday:2009aq, Wyllard:2009hg} concerns the specific example where $k$ is unity.  In that case the coset models appearing in \eqref{duality2} have no degrees of freedom and reduce to ordinary real Toda theory.  According to \cite{Forgacs:1989ac}, the real Toda theory is related to the $\frak{sl}(n,\mathbb{R})$ Chern-Simons theory in the same way that the  $\frak{sl}(n,\mathbb{C})$  Chern-Simons theory is related to the complex Toda theory.  Therefore when $k=1$ we can gain evidence for \eqref{duality2} by directly relating $\frak{sl}(n,\mathbb{R})$ and $\frak{sl}(n,\mathbb{C})$  Chern-Simons theory.

Geometric quantization of $\frak{sl}(n,\mathbb{R})$ and $\frak{sl}(n,\mathbb{C})$  Chern-Simons theory has been investigated in detail in \cite{Dimofte:2011gm, Dimofte:2014zga} and indeed when $k=1$ there is a close connection between their quantization.  As a specific example we consider, following \cite{Dimofte:2011py, Dimofte:2014zga}, the phase space
\begin{equation}
\mathbb{C}^{*}\times \mathbb{C}^{*}=\{(\mathcal{X}, \mathcal{Y})\}~, \hspace{.5in} \Omega = \frac{d\mathcal{X}}{\mathcal{X}}\wedge \frac{d\mathcal{Y}}{\mathcal{Y}}~, \label{phase}
\end{equation}
where $\Omega$ is a holomorphic symplectic form.  This phase space is relevant to complex Chern-Simons theory because it occurs as the space of flat $\frak{sl}(2,\mathbb{C})$ connections on sphere with four-punctures with unipotent holonomy at each puncture.  A wide variety of larger phase spaces may be built by appropriately gluing this basic model \cite{Dimofte:2011gm, Dimofte:2013iv}.

Complex Chern-Simons quantizes this phase space using the real symplectic form
\begin{equation}
\omega=\frac{(k+is)}{4\pi}\Omega+\frac{(k-is)}{4\pi}\bar{\Omega}~.\label{realsymp}
\end{equation}
As in our discussion of the canonical variables \eqref{canonicalvars} for complex Liouville theory, the physics of \eqref{realsymp} is most clear in a basis of fields where the syplectic form is diagonal.  Such a basis is easily found resulting in
\begin{equation}
\omega =\frac{2\pi}{k}\left(d\nu\wedge d\mu-dn\wedge dm\right)~.
\end{equation}
The variables $\mu, \nu, m, n$ are all real.  Meanwhile, $\mu$ and $\nu$ are continuous while $m$ and $n$ are periodic and identified mod $k$.

Canonical quantization now promotes these variables to operators obeying
\begin{equation}
[\mu,\nu]=-\frac{k}{2\pi i}~, \hspace{.5in}[m,n]=\frac{k}{2\pi i}~.
\end{equation}
Note that since $m$ and $n$ are both periodic, their spectrum is quantized.  Therefore the Hilbert space associated to these variables is finite dimensional.  In total then, the quantization of \eqref{phase} produces
\begin{equation}
L^{2}(\mathbb{R})\otimes \mathbb{C}^{k}~.
\end{equation}
In particular, when $k$ is unity the finite-dimensional space of degrees of freedom drops out and we are left with simply $L^{2}(\mathbb{R})$.  This is exactly the expected Hilbert space from quantizing $\frak{sl}(2,\mathbb{R})$ Chern-Simons theory on the same manifold and supports our claim of a direct relationship between these two theories when $k=1$.

\subsubsection{The Case $k=2$}

The case $k=2$ has been investigated in \cite{Belavin:2011sw, Schomerus:2012se} (where the imaginary part of $\Phi$ was referred to as a ``shadow Liouville" field to account for its unusual contour of integration).

In this case the duality relates complex Liouville to $\mathcal{N}=1$ superLiouville plus the decoupled coset $\hat{\frak{su}}(2)_{2}/\hat{\frak{u}}(1)$.  This coset model is equivalent to a free fermion.  Therefore all told the model contains two fermions; one interacting and one free and we assemble them into a single complex fermion $\chi$.  To relate to complex Liouville theory one then proceeds by bosonization.  For $j\in \mathbb{Z}$ we introduce exponential vertex operators defined using a periodic scalar $\theta$\footnote{For half-integral $j$ analogous expressions exist involving the fermionic spin field.}
\begin{equation}
:\exp(i j \theta):=\frac{1}{j!}:\prod_{i=0}^{j-1}\partial^{i}\chi\bar{\partial}^{i}\bar{\chi}:~. \label{vertextheta}
\end{equation}
The vertex operators of the theory then are expressed in terms of the bosonic Liouville field $\phi$ and \eqref{vertextheta}.
\begin{equation}
\mathcal{O}_{\alpha,j}=:\exp(\alpha \phi+ij\theta):~. \label{odef}
\end{equation}
Note that since $\phi$ and $\chi$ interact the vertex operators $\mathcal{O}_{\alpha,j}$ are not simply products of the $\phi$ vertex operators and the $\theta$ vertex operators \eqref{vertextheta}.

So far \eqref{odef} is just a definition.  To see the relationship with complex Liouville, \cite{Belavin:2011sw, Schomerus:2012se} factorized the vertex operators as
\begin{equation}
\mathcal{O}_{\alpha,j}(z,\bar{z})=V(z,\bar{z})\tilde{V}(z,\bar{z})~.
\end{equation}
Here $V$ and $\tilde{V}$ are vertex operators in two separate Liouville theories.  Moreover, as  \cite{Belavin:2011sw, Schomerus:2012se} verifeid, the three point functions of super-Liouville plus the free fermion respect this splitting.  However, the momenta of these vertex operators are correlated.  It is natural to interpret these two vertex operators as those associated to the $\Phi$ and $\bar{\Phi}$ fields of \eqref{clvl} lending further support to our conjectured duality.

\section{Half-BPS Operators and W-Algebras}
\label{sec:operators}

We may make use of our construction to understand a variety of similar relationships between the Toda theories and the $(2,0)$ theories.  The relationships described here provide direct information about properties of the $(2,0)$ theory in flat space.  Throughout, we assume that the $k=1$ version of the duality discussed in the previous section holds.

\subsection{The SuperConfomal Index and W-Characters}

A basic data of the $(2,0)$ theory is its spectrum of local operators.  These operators fall into unitary representations of the $(2,0)$ superconformal algebra $\frak{osp}(8|4)$.  The smallest non-trivial representations of this algebra are half-BPS \cite{Minwalla:1997ka}.  Their superconformal primary is a Lorentz scalar transforming in a traceless symmetric tensor representation of the $\frak{so}(5)_{R}$ symmetry.  They obey the shortening condition
\begin{equation}
Q_{\alpha}^{(i}\mathcal{O}^{A_{1} \cdots A_{n})}=0~, \hspace{.5in}\Delta(\mathcal{O})=2n~, \label{halfbpsops}
\end{equation}
where in the above the symmetrization denotes the projection onto the irreducible $\frak{so}(5)_{R}$ representation with largest dimension.  A first pass at understanding the $(2,0)$ theories is to describe the spectrum of these operators in detail.

This question has been addressed from a variety of perspectives.  To begin, one may note that the moduli space of the six-dimensional $(2,0)$ of type $\frak{g}$ has a moduli space $(\mathbb{R}^{5})^{\mathrm{rank}(\frak{g})}/S_{\frak{g}},$ where $S_{\frak{g}}$ is the Weyl group.  One may anticipate that the generators of the algebra of functions on this moduli space is exactly this set of half-BPS operators \cite{Bhattacharyya:2007sa}.  Therefore the half-BPS operators consist of the Casimir invariants of the algebra $\frak{g}$.  Thus there are $\mathrm{rank}(\frak{g})$ multiplets of the type \eqref{halfbpsops}, with the scaling dimension of the $i$-th operator equal to twice the degree $d_{i}$ of the associated Casimir invariant.\footnote{Thus for $\frak{g}=A_{n-1}$ the $d_{i}$ are $2, 3, \cdots, n.$}  This result may also be produced from light-cone quantization descriptions of the $(2,0)$ theory \cite{Aharony:1997th, Aharony:1997an}.

One consequence of this spectrum of half-BPS operators is a simple formula for a limit of the superconformal index \cite{Kim:2012ava, Kim:2013nva}.  This limit counts the half-BPS multiplets (as well their products and some derivatives).  It takes the form
\begin{equation}
\mathcal{I}(q)=\mathrm{Tr}\left[(-1)^{F}q^{\Delta-R}\right]=\prod_{i=1}^{\mathrm{rank}(\frak{g})} \prod_{n=0}^{\infty}\frac{1}{1-q^{n+d_{i}}}~, \label{Wcharacter}
\end{equation}
where the trace is over the space of local operators, and $R$ is a Cartan generator of $\frak{so}(5)_{R}$ that counts the number of vector indices.

An intriguing observation of \cite{Beem:2014kka} is that \eqref{Wcharacter} is equal to the vacuum character of a two-dimensional chiral $\mathcal{W}_{\frak{g}}$ algebra.  This is the two-dimensional vertex operator algebra with primaries given by holomorphic currents $\chi_{i}$ with scaling dimensions $d_{i}$.  In fact, as argued in \cite{Beem:2014kka}, the relationship between the chiral $\mathcal{W}_{\frak{g}}$ algebra and the half-BPS operators of the (2,0) theory runs deeper than a bijection of sets.  Indeed, the half-BPS operators (suitably twisted) have a closed operator algebra.  Therefore it is natural to conjecture that this operator algebra is exactly the $\mathcal{W}_{\frak{g}}$ algebra.

We will be able to derive the result \eqref{Wcharacter} and some aspects of the stronger claims of \cite{Beem:2014kka} in a simple way using the techniques developed in the previous sections.  To begin, recall that the chiral $\mathcal{W}_{\frak{g}}$ algebra is exactly the chiral sector of the real Toda field theory.  It is easy to see this using the relation to the WZW model described in the previous sections.  Before imposing constraints, the dimensions of the holomorphic currents are all unity.  After imposing the constraints, the Virasoro generator is shifted as in \eqref{modifiedl0} to $L_{0}'=L_{0}-H$.  The currents which remain after the constraints are imposed have exactly the scaling dimensions, with respect to $L_{0}',$ given by the degrees of the $d_{i}$ of the Weyl invariants.

Therefore, we aim to show that the chiral algebra operators contributing to the index \eqref{Wcharacter} are exactly the chiral algebra of the Toda theory.  To derive this, we view the operators as BPS states on $S^{5}$ via the state operator correspondence.  The index, \eqref{Wcharacter} is an  $S^5 \times S^1$ partition function with a Wilson line in the Cartan of the $\frak{so}(5)_{R}$ symmetry and no chemical potentials for the rotations (in other words, the space is metrically a direct product of $S^{5}$ and $S^{1}$). This background may be conformally mapped to one of the type studied here, with the same type of boundary conditions.

As in the previous sections, the key idea is to find a suitable $S^{3}$ hiding in the geometry and to reduce the problem to a question of edge modes in complex Chern-Simons theory.  This is achieved by viewing $S^{5}$ as an $S^3$ fibered over a disk, where the radius of the $S^3$ vanishes at the edge of the disk. The boundary conditions in the 5d Yang-Mills language, after Hopf reduction in the $S^3$, will again be of the form of D4 branes ending on a single D6 brane, namely the maximal Nahm pole. Therefore we can attempt again to reduce on the $S^{3}$ and describe the resulting states as edge modes in Chern-Simons theory.

The only thing we need to check is that there is a Weyl rescaling that relates the index background to that of the 3d-3d correspondence. It this case, it is almost obvious. The Lorentz signature background $S^5 \times \mathbb{R}^1$ is conformally flat, thus it is conformally equivalent to $S^3 \times \textrm{AdS}_3$ with equal radii. The Weyl transformation is divergent along an equator of $S^5$ ($\times \mathbb{R}^1$), which maps to the boundary of AdS$_3$.  Furthermore, the supersymmetric background $S^3 \times M_3$ found in \cite{Cordova:2013cea} has the property that if $M_3$ is AdS$_3$ with the same radius as $S^3$, the $R$ gauge fields along $M_3$ in the $SO(3) \subset SO(5)$ vanish. This is due to a cancellation between the $R$ gauge fields required for supersymmetry on $S^3$ with the usual twisting in the curved 3-manifold, AdS$_3$.

Therefore, in summary, the $(2,0)$ theory, reduced on $S^{5}$ has states described by complex Chern-Simons on AdS$_{3}$.  This is toplogically the disc $\times$ time.  Since the $S^{3}$ shrinks at the boundary, we again obtain the Nahm pole boundary conditions carrying out the Hamiltonian reduction of the WZW model.  The only difference from the AGT setup is that in this case, the theory is chiral since now we have only a single boundary component.  Therefore we obtain the chiral constrained WZW model.  Moreover, in this setup, the discrete level $k$ is one, and the parameter $b$ is unity.  So the additional coset models are trivial and we find exactly the $\cal{W}_{\frak{g}}$ chiral algebra states.  In particular we obtain the $S^{5}\times S^{1}$ partition function \eqref{Wcharacter}.

One important consistency check on this result, concerns the dependence on the squashing parameter $\ell$.  In our setup, this changes the $b$ parameter of the chiral $\mathcal{W}_{\frak{g}}$ algebra and hence changes the chiral correlators for instance by deforming the central charge.  However the torus partition function of the $\mathcal{W}_{\frak{g}}$ algebra does not depend continuously on this parameter and instead produces the partition function \eqref{Wcharacter} even after deforming $b$ away from unity.  Since the squashing parameter is deforms the ratio of radii of the $S^{2}$ and $S^{1}$ in the Hopf fibration of $S^{3},$ it may be viewed in the superconformal index setup as an additional chemical potential for rotations.  Therefore our setup predicts that the superconformal index of the $(2,0)$ theory does not depend on one chemical potential for rotations.

Happily, this prediction is also known to be correct.  Indeed, one may generalize the limit of the superconformal index as
\begin{equation}
\mathcal{I}(q,s)=\mathrm{Tr}\left[(-1)^{F}q^{\Delta-R}\lambda^{2j_{1}}\right]~,
\end{equation}
where $j_{1}$ is a rotation quantum number.  Although it is in principle possible that the index could depend on $\lambda$ it in fact is known that it does not \cite{Kim:2012ava, Kim:2013nva}, and the answer reduces to \eqref{Wcharacter} for all $\lambda$ as expected.

\subsection{W-Algebras and Conformal Anomalies}

As a final application we shed light on a connection between the correlators of operators in the chiral algebra of the $(2,0)$ theory and $\mathcal{W}_{\frak{g}}$ algebras conjectured in \cite{Beem:2014kka}.  In particular this explains a relation between the conformal anomalies of the $(2,0)$ theory and that of chiral Toda.

Let us begin with a brief review of the conformal anomalies.  In general in six-dimensional conformal field theory coupled to a background metric the trace of the energy momentum tensor has an anomolous expectation value
\begin{equation}
\langle T^{A}_{A}\rangle= a E_{6} +c_{1}I_{1}+c_{2}I_{2}+c_{3}I_{3}~,
\end{equation}
where $E_{6}$ is the Euler density of the background metric and $I_{i}$ are invariants constructed from the Weyl tensor (see e.g. \cite{Bastianelli:2000hi} for precise expresions.)  The numbers $a$ and $c_{i}$ are conformal anomalies.  They may also be defined by various pieces of correlation functions of the energy momentum tensor in flat space.  For instance $a$ first occurs in a four-point function while the $c_{i}$ occur in two and three-point functions.

In the context of $(2,0)$ supersymmetry it is known that the $c_{i}$ are not independent, and in fact are all equal.  Therefore we write $c_{i}=c$ and fix conventions such that $c_{i}$ of the free $(2,0)$ tensor multiplet is one.  The $c$ anomaly for general (2,0) theory of type $\frak{g}$ is then given by
\begin{equation}
c_{\frak{g}}=4h^{\vee}d_{\frak{g}}+r_{\frak{g}}~.  \label{canomaly}
\end{equation}
For $\frak{g}=\frak{su}(n)$ this shows the characteristic $n^{3}$ behavior at large $n$ first found holographically \cite{Henningson:1998gx}.  The finite $n$ corrections have been studied holographically in \cite{Tseytlin:2000sf, Beccaria:2014qea}, and also determined independently from the point of view of anomaly multiplets \cite{Cordova:2015vwa, Cordova:2015fha, Beccaria:2015ypa}.

For our purposes, the most striking aspect of \eqref{canomaly} is that it is exactly equal to the central charge of the chiral Toda theory at $b=1$.  This observation was one of the original pieces of evidence of the AGT conjecture and its orbifold generalizations \cite{Alday:2009qq, Nishioka:2011jk}.

This observation plays a natural role in the setup of \cite{Beem:2014kka}.  In that context one of the chiral operators contributing to the index \eqref{Wcharacter} is the $6d$ energy-momentum tensor, and in the relation to the $\cal{W}_{\frak{g}}$ algebra, this operator maps to the dimension two Virasoro current.  Part of the conjecture of \cite{Beem:2014kka} is that ($R$-twisted) correlators of the $6d$ chiral operators must agree with those of the $\cal{W}_{\frak{g}}$ algebra hence in particular the two point function of stress tensors in $6d$ and in $2d$ must agree.  Since the two point functions are controlled by the $c$-anomaly \eqref{canomaly} in $6d$ and the central charge in $2d$, this implies as a corollary the relation between the conformal anomalies.

We can motivate these results in a straightforward way by a suitable compactification and reduction.  We consider the $(2,0)$ theory on $S^{6}_{\ell}$, where the special case $\ell=1$ is the round sphere.  This space is topologically a $S^{6}$ but is metrically deformed.  It is conveniently viewed as a fibration of $S^{3}_{\ell}$ over a solid ball.  Explicitly, the metric is
\begin{equation}
ds^{2}=r^{2}\left[d\sigma^{2}+\phantom{\frac{1}{2}}\hspace{-.15in}\sin(\sigma)^{2}d\Omega_{2}^{2}\right]+   r^{2}\left[\frac{\cos(\sigma)^{2}\ell^{2}}{4} \left(d\theta^{2}+\sin^{2}(\theta)d\phi^{2}\right) +\cos(\sigma)^{2}\left(d\psi+\cos^{2}(\theta/2)d\phi\right)^{2}\right]~,
\end{equation}
where $d\Omega_{2}^{2}$ is the round metric on a $S^{2}$ of unit radius.  The factors in brackets are the metric on the ball and on $S^{3}_{\ell}$.  At the boundary of the ball $S^{3}_{\ell}$ shrinks.

The $6d$ (2,0) theory may be placed on $S^{6}_{\ell}$ preserving supersymmetry.  As usual this can be seen by a conformal transformation to make the $S^{3}_{\ell}$ have constant size, which again maps the background to a special case of the $3d$-$3d$ correspondence.  We can add interesting observables to this construction by placing local operators from the chiral algebra, in particular the half-BPS operators, at points on the $S^{2}$ at the edge of the ball when $\sigma=\pi/2$.   In particular, in the case $\ell=1$ the geometry is conformally flat and these correlators should reproduce the flat space chiral algebra of \cite{Beem:2014kka}.

We claim that these correlators are exactly those of the chiral $\cal{W}_{\frak{g}}$ algebra with $b$ related to the squashing parameter in the usual way
\begin{equation}
\frac{b+b^{-1}}{2}=\ell^{-1}~.
\end{equation}
To see this in our setup we simply proceed along the usual lines and reduce along the $S^{3}_{\ell}$.  This results, again, in complex $\frak{g}_{\mathbb{C}}$ Chern-Simons theory on the solid ball.  Moreover since the $S^{3}_{\ell}$ again collapses at the edge of the ball, we obtain the Nahm pole boundary conditions as before.  Therefore, the edge mode theory is again the chiral Toda theory this time on the $S^{2}$.

To see the connection with the correlators, we must now argue that the insertions of half-BPS operators in the $(2,0)$ theory map to insertion of the  $\cal{W}_{\frak{g}}$ algebra currents in the boundary.  This can be seen using the reduction to five-dimensions and our knowledge of the Chern-Simons fields.  The superconformal primaries of the half-BPS operators descend to Casimir invariants of the scalars, i.e. expressions of the form $\mathrm{Tr}(X^{k})$.  These are exactly the currents of the $\cal{W}_{\frak{g}}$ algebra.  Indeed using the analysis of section \ref{sec:fluc}, we see that the holomorphic current $W_{j}(u)$ of dimension $j$ is given by the scalar $X_{u}$ with
\begin{equation}
X_{u}\sim T_{+}+W_{j}(u)T_{-}^{j}~,
\end{equation}
with $T_{-}$ the lowering generator picked out by the Nahm pole.  Taking traces of powers, we then $W_{j}$.  Therefore correlators of half-BPS operators of the $(2,0)$ theory, restricted to a plane, i.e. the $S^{2}$ in our geometry, must agree with $\cal{W}_{\frak{g}}$ algebra correlators.  In particular, the central charges must agree.

\section*{Acknowledgements}
We are grateful to C. Beem, T. Dimofte, T. Dumitrescu, L. Rastelli, C. Vafa, E. Witten, and X. Yin for discussions. CC is supported by a Schmidt fellowship at the Institute for Advanced Study, and DOE grant DE-SC0009988.  D.L.J. is supported in part by NSF grant PHY-1352084 and a Sloan Fellowship.

\bibliography{AGT}{}
\bibliographystyle{utphys}

\end{document}